\documentclass[aps,prl,amsmath,amssymb,superscriptaddress,showkeys,showpacs,twocolumn,floatfix]{revtex4-2}
\usepackage[final]{graphicx}
\usepackage{hyperref}
\usepackage{amsmath}
\usepackage{bbm}
\usepackage{amsfonts}
\usepackage{amssymb}
\usepackage{latexsym}
\usepackage{graphicx}
\usepackage[english]{babel}
\usepackage{multirow}
\usepackage{float}
\usepackage{url}
\usepackage{slashed}
\usepackage{xcolor}
\usepackage[utf8]{inputenc}
\usepackage{verbatim}
\usepackage{lipsum}

\usepackage[utf8]{inputenc}

\usepackage{diagbox}

\usepackage{mathtools}
\usepackage{amsfonts}
\usepackage{mathrsfs}
\usepackage{bbm}
\usepackage{slashed}

\usepackage{graphicx}
\usepackage{color}
\usepackage{array}
\usepackage{esint}
\usepackage{placeins}
\usepackage{booktabs}
\usepackage{makecell}
\usepackage{epstopdf}
\usepackage[caption=false]{subfig}

\usepackage{xspace}
\usepackage{siunitx}
\usepackage{hyperref}
\usepackage[nameinlink]{cleveref}
\usepackage{appendix}

\usepackage{xifthen}
\usepackage{xcolor}
\hypersetup{
	colorlinks,
	linkcolor={red!75!black},
	citecolor={blue!75!black},
	urlcolor={blue!75!black}
}
\usepackage{comment}

\renewcommand{\d}{\mathrm{d}}

\begin{document}

\title{Vacuum Decay Rate in D-dimensional Electroweak theories}
\author{Jingwei Wang}
\affiliation{Department of Physics and Chongqing Key Laboratory for Strongly Coupled Physics, Chongqing University, Chongqing 401331, P. R. China}

\author{Ligong Bian\footnote{Corresponding Author.}}
\email{lgbycl@cqu.edu.cn}
\affiliation{Department of Physics and Chongqing Key Laboratory for Strongly Coupled Physics, Chongqing University, Chongqing 401331, P. R. China}

\date{\today}

\begin{abstract}
We present a systematic  framework for computing the functional determinant contribution to vacuum decay rates in D-dimensional electroweak theories. It consistently incorporates quantum fluctuations from scalar, fermion, and gauge fields while ensuring rapid convergence even at high angular momenta. 
Demonstrated through applications to the 
$D=4$ SMEFT and its  
$D=3$ thermal counterpart, our method provides a general and efficient tool for analyzing vacuum stability and decay across dimensions in the presence of new physics beyond the Standard Model.
\end{abstract}
\maketitle

\noindent\textit{Introduction--}
The stability of the electroweak vacuum is a fundamental question in particle physics and cosmology. The current Higgs potential may reside in a metastable rather than an absolutely stable state at extremely high energy scales, suggesting that our universe might merely exist within a long-lived but not eternal quantum state. This ``metastability" offers a new perspective for addressing the hierarchy problem in theoretical physics~\cite{Branchina:2013jra,Espinosa:2015qea,Degrassi:2012ry,Casas:1994qy,Branchina:2013jra,Casas:1996aq,Cankocak:2020pqd,Bentivegna:2017qry,Casas:1996aq,Branchina:2014usa,Greenwood:2008qp}. Its decay rate is directly connected to grand unification theories~\cite{Branchina:2014rva},  inflation~\cite{Fumagalli:2019ohr,Shkerin:2015exa,Herranen:2015ima,Herranen:2014cua}, black holes~\cite{Espinosa:2017sgp,Hayashi:2020ocn,Sengupta:2025jah} and gravitational waves~\cite{Espinosa:2018eve},  pointing toward the ultimate fate of the universe. 

The standard procedure for calculating the vacuum decay rate can be found in Refs.~\cite{colemanFateFalseVacuum1977,callanFateFalseVacuum1977} . This is a semiclassical approach based on the bounce solution, which expresses the vacuum decay rate as $\gamma=\mathcal{A}e^{-\mathcal{B}}$, where $\mathcal{A}$ and $\mathcal{B}$ are the prefactor arising from quantum corrections and the action of a bounce solution, respectively. 
However, in general — and also in this work — the leading contribution to $\mathcal{A}$ comes from the functional determinant which also plays an important role in thermal bubble nucleation~\cite{Linde:1981zj,Linde:1980tt}. Neglecting higher-order corrections, we will therefore simply take $\mathcal{A}$ to denote this one-loop functional determinant in what follows.
Regarding the computation of the decay rate, most papers focus on $\mathcal{B}$, while the remaining ones primarily concentrate on scalar fluctuations (such as Refs.~\cite{PhysRevD.91.105021,baackeOneloopCorrectionsMetastable2004,Dunne:2005rt}), partly due to the challenges in calculating contributions from fermions and bosons. In terms of computational methods, Ref.~\cite{chigusaDecayRateElectroweak2018} which calculated the functional determinant of  vacuum decay rate in the Standard Model actually provides a method for the four-dimensional case. 
Ref.~\cite{baratellaFunctionalDeterminantsFalse2025} offers a low-order approximation for the contributions of various fields to $\mathcal{A}$ in four dimensions, while Ref.~\cite{dunneFunctionalDeterminantsRadial2006,k1b6-2gbs,Ekstedt:2023sqc} provide higher-order approximations for the contributions of the Higgs field to $\mathcal{A}$ in $D$-dimensional spacetime.

In this Letter, we present a convergent and dimensionally consistent framework for computing the functional determinant of vacuum decay rates in $D$-dimensional electroweak theories, achieved by combining a WKB expansion with dimensional regularization. The scheme systematically incorporates scalar, fermion, and gauge field fluctuations, overcoming the convergence issues and low-order limitations that have constrained previous calculations, and enabling arbitrary‑order WKB approximations in any dimension $D$. Going beyond the Standard Model vacuum lifetime studies~\cite{Isidori:2001bm,chigusaDecayRateElectroweak2018,Chigusa:2017dux,Andreassen:2017rzq,Baratella:2024hju}, our method applies to false vacuum decay driven by new physics and to early-universe symmetry-breaking transitions. We illustrate its utility through concrete applications to the $D=4$ and dimensionally reduced $D=3$ formulations of the Standard Model effective field theory (SMEFT). 
To the best of our knowledge, previous studies of the nucleation rate of first-order electroweak phase transition (EWPT) have largely focused on scalar fields only (e.g. Refs.~\cite{Munster:1999hr,Dunne:2005rt,Baacke:1993ne,Baacke:2003uw,Baacke:2008zx,Garbrecht:2015oea}) or the 4D studies at zero temperature (see Refs.~\cite{Ai:2018guc,Ai:2023yce,Ai:2020sru} for examples). Therefore, this study constitutes a theoretical calculation for such a first-order EWPT that is the first to completely include the functional determinants of gauge bosons and integrated-out fermions in three dimensions.

\noindent\textit{Set up--}
Since  the functional determinant $\mathcal{A}$ is only related to the second order Euclidean action, the non-Abelian part of the gauge field kinetic term in the SU(2) model enters through the group volume factor when handling gauge zero modes. Therefore, the difference between the SU(2) model and the U(1) model lies solely in the volume of the group space here. 

We work in the Fermi gauge, 
because it allows us to neglect the contribution from ghost fields \cite{kusenkoVacuumDecayInternal1997} and avoids the kind of troubles described in Ref.~\cite{endoGaugeInvarianceDecay2017}. As a nontrivial check, the gauge parameter $\xi$ explicitly cancels between $\det \Psi_\nu$ and $\det \hat{\Psi}_\nu$ in the Da mode, ensuring the final result is independent of $\xi$. The explicit verification is presented in the \textit{Supplemental Material}. For more general gauge scenarios, we left to future work. We start directly from the second-order term of the D-dimensional spacetime electroweak Euclidean action
\begin{equation}
	S_2=\frac{R^{D-2l-2}}{\kappa^2}\int \d^Dx \left(\mathcal{L}_{\text{H}}+\sum\mathcal{L}_{\psi}+ \sum\mathcal{L}_{A_\mu,a}\right),
\end{equation}
With $\bar{\phi}$, $h$, $\psi$, $A_\mu$, $a$ and $c$ represent the (re-scaled) bounce solution , Higgs field, fermion field, gauge field, NG boson field and ghost field, respectively, and denote $X_y\equiv yX$, $X_g\equiv gX$, $m^2_\phi=\frac{\d^2 U}{\d \bar{\phi}^2}$ and $m^2_a=\left. \frac{\d^2 U}{\d a^2}\right|_{a=0} +\bar{\phi}^2_g$, where $y$ and $g$ are the corresponding coupling constants originating from the Yukawa interaction term $y\bar{\phi}\bar{\psi}\psi$ and the covariant derivative term $D_\mu=\partial_\mu-ig^aA_\mu^a \tau^a$ with $g^a=\left(g,g,g,g'\right)$, $A^a_\mu=\left(A^1_\mu,A^2_\mu,A^3_\mu,B_\mu\right)$ and $\tau^a=\left(\sigma^1,\sigma^2,\sigma^3,I_2\right)$, the Euclidean Lagrangian of the $X$ field are:
\begin{equation}
	\begin{aligned}
		\mathcal{L}_{\text{H}}=&\frac{1}{2}h\left(-\partial^2+m^2_\phi\right)h,\\
		\mathcal{L}_{\psi}=&\bar{\psi}\left(\slashed{\partial}+\bar{\phi}_y\right)\psi,\\
		\mathcal{L}_{A_\mu,a}=&\frac{1}{2}A_{\mu}\left[-\partial^2+\left(1-\frac{1}{\xi}\right)\partial_\mu\partial_\nu+\bar{\phi}^2_g\right]A_{\mu}\\
		&+2\left(\partial_\mu \bar{\phi}_g\right)A_\mu a+\bar{\phi}_ga\partial_\mu A_\mu\\
		&+\frac{1}{2}a\left(-\partial^2+m^2_a-\bar{\phi}^2_g\right)a-\bar{c}\partial^2c\;.
	\end{aligned}
\end{equation}

Introducing the re-scaling constants of coordinates, couplings and fields: $R$, $\kappa$ and $l$, the bounce action $\mathcal{B}$ can be simply written as
\begin{equation}
	\mathcal{B}=\frac{\Omega_DR^{D-2l-2}}{\kappa^2}\int_0^{+\infty}  r^{D-1}\left[\frac{1}{2}\left(\bar{\phi}^\prime\right)^2+U(\bar{\phi})\right]\d r,
\end{equation}
where $r=\sqrt{x_\mu x_\mu}$, $\Omega_D$ is the area of a D-dimensional unit sphere,  and here and hereafter, both $\partial_r X$ and $X^\prime$ represent $\frac{\d X}{\d r}$, with the difference that $X^\prime$ acts only on a single object $X$, while $\partial_r$ will act on all subsequent objects as much as possible.
It can be seen that since the entire theory is developed around the bounce solution $\bar{\phi}$, the mathematical properties of $\bar{\phi}$ are crucial. We define $\Delta_\nu$ as the following operator
\begin{equation}
	\Delta_\nu=\partial_r^2+\frac{D-1}{r}\partial_r-\frac{L^2}{r^2},
\end{equation}
where $L^2=\left(\nu-1\right)\left(\nu+D-3\right)$ and $\nu=1,2,3,\cdots$. The equation governing the evolution of $\bar{\phi}$ can be expressed as \cite{colemanFateFalseVacuum1977}
\begin{equation}\label{equ-8}
	\Delta_1\bar{\phi}=\frac{\d U}{\d\bar{\phi}},
\end{equation}
subject to the boundary conditions $\bar{\phi}^\prime(0)=0$ and $\bar{\phi}(r\rightarrow\infty)=v$. Here the center of the bounce is set to $r=0$. 

Since the equation is singular at $r=0$, we need to perform a Taylor expansion in its vicinity to avoid the singularity. We obtain
\begin{equation}
	\left. \bar{\phi}\right|_{r\rightarrow+0} =\phi_c + \phi_2r^2 + \dots
\end{equation}
where $\phi_2=\left.\frac{1}{2D}\frac{\d U}{\d \bar{\phi}}\right|_{\bar{\phi}=\phi_c}$.
On the other hand, the asymptotic solutions of the bounce solution at infinity can be expressed as
\begin{equation}\label{equ-11}
	\bar{\phi}\simeq v+ \phi_\infty K_{\frac{D}{2}-1}(\hat{m}_\phi r)\left(\frac{\hat{m}_\phi}{r}\right)^{\frac{D}{2}-1}
\end{equation}
for $\hat{m}_\phi>0$, where $K_{\frac{D}{2}-1}(x)$ is the modified Bessel function of the second kind and $\phi_\infty$ is a constant. Here and after, we denote by $\hat{X}$ the counterpart of  $X$ in the false vacuum. And it is
\begin{equation}
	\bar{\phi}\simeq v+ \phi_\infty\frac{2^{\frac{D}{2}-2}}{r^{D-2}}\Gamma(\frac{D}{2}-1)
\end{equation}
for $\hat{m}_\phi=0$, where we assume $D>2$, and $\Gamma(x)$ is the Gamma function. This conclusion can also be reached by taking the limit $\hat{m}_\phi\rightarrow+0$ of \eqref{equ-11}\footnote{ Due to the complexity of the relevant scenarios, the cases where $\hat{m}_\phi=0$ when $D=2$ 
are not considered in this Letter.}.

\noindent\textit{Prefactors and Functional Determinants--}
At the second order of action in the fluctuations around the bounce, the prefactor can be decomposed as
\begin{equation}
	\mathcal{A}=\mathcal{V}_D^{-1}\mathcal{A}_{\text{H}}^{-\frac{1}{2}}\prod\mathcal{A}_{\psi}\prod\mathcal{A}_{A_\mu,a}^{-\frac{1}{2}},
\end{equation}
where $\mathcal{V}_D$ is the volume of spacetime. Meanwhile, $\mathcal{A}_{X}$ represents the contribution of particle X. 
%
Subsequently, we should employ the $O(D)$ symmetry of the bounce solution to expand the determinant and utilize the Gelfand-Yaglom theorem \cite{gelfandIntegrationFunctionalSpaces1960,dashenNonperturbativeMethodsExtendedhadron1974,formanFunctionalDeterminantsGeometry1987,KirstenMathematicalSciencesResearchInstitute2010,Kirsten_2004,KIRSTEN2003502}. Ultimately, we obtain a result of the form
\begin{equation}\label{equ-16}
	\ln \mathcal{A}_{X}=\sum_{\nu=1}^{+\infty}\text{deg}_X(\nu;D)\ln \frac{\Psi_\nu^X}{\hat{\Psi}_\nu^X},
\end{equation}
where $\text{deg}_X(\nu;D)$ is the degeneracy obtained by expanding the prefactor in terms of the basis composed of D-dimensional spherical harmonics, while $\Psi^{X}_\nu=\lim\limits_{r\rightarrow+\infty}\Psi^{X}_\nu(r)$
and $\hat{\Psi}_\nu^X=\lim\limits_{r\rightarrow+\infty}\hat{\Psi}^{X}_\nu(r)$. Originally, $\Psi^{X}_\nu(r)$ and $\hat{\Psi}^{X}_\nu(r)$ were the functional determinants  of the operator matrices, with $\lim\limits_{r\rightarrow+0}\frac{\Psi^{X}_\nu}{\hat{\Psi}_\nu^X}=1$. But here we have computed the result of the determinants and obtained individual scalar functions along with the equations they satisfy.
$\text{deg}_X(\nu;D)$ are respectively as follows:
\begin{enumerate}
	\item Higgs
	\begin{equation}
		\text{deg}_\text{H}(\nu;D)=\frac{\left(D+2\nu-4\right)\Gamma(D+\nu-3)}{\Gamma(D-1)\Gamma(\nu)},
	\end{equation}
	\item Fermions
	\begin{equation}
		\text{deg}_\psi(\nu;D)=2^{\lfloor\frac{D}{2}\rfloor}\frac{\Gamma(\nu+D-2)}{\Gamma(\nu)\Gamma(D-1)},
	\end{equation}
	\item The `diagonal' modes of gauge boson $A_\mu$ mixed with the NG boson $a$ (denoted `Da modes')
	\begin{equation}
		\text{deg}_\text{Da}(\nu;D)=\text{deg}_H(\nu;D),
	\end{equation}
	\item The other contains the transverse modes of $A_\mu$ (denoted `T modes')
	\begin{equation}
		\text{deg}_\text{T}(\nu;D)=\frac{\left(D+2\nu-2\right)\left(D+\nu-2\right)\Gamma(D+\nu-3)}{\left(\nu+1\right)\Gamma(D-2)\Gamma(\nu)}.
	\end{equation}
\end{enumerate}
Here, $\lfloor X\rfloor$ is the floor function. It can be seen that these results agree with the conclusions in other papers in four dimensions, such as Ref.~\cite{chigusaDecayRateElectroweak2018} for Higgs and fermion parts, and Ref.~\cite{baratellaRevisingFullOneLoop2025} for the gauge contributions. 
And, $\Psi^{X}_\nu(r)$'s satisfy the following equations:
\begin{enumerate}
	\item Higgs ($\nu\neq2$)
	\begin{equation}\label{equ-24}
		\left(\Delta_\nu-m^2_\phi\right) \Psi^{\text{H}}_\nu(r)=0,\\
	\end{equation}
	\item Fermion
	\begin{equation}\label{equ-25}
		\left(\Delta_\nu-\bar{\phi}^2_y\right) \Psi^{\psi}_\nu(r)=\frac{\bar{\phi}_y^\prime}{\bar{\phi}_y}\left(\partial_r-\frac{\nu-1}{r}\right)\Psi^{\psi}_\nu(r),
	\end{equation}
	\item Da mode ($\nu>1$)
	\begin{equation}\label{equ-26}
		\left( \Delta_\nu-m^2_a\right)  \Psi_\nu^{\text{Da}}(r)=\frac{\partial_r\left(\bar{\phi}^2_g r^2\right)}{\bar{\phi}^2_g r^2+L^2}\left(\partial_r-\frac{\bar{\phi}_g^\prime}{\bar{\phi}_g}\right)\Psi_\nu^{\text{Da}}(r),
	\end{equation}
	\item T mode
	\begin{equation}\label{equ-27}
		\left( \Delta_{\nu+1}-\bar{\phi}^2_g\right)\Psi_\nu^{\text{T}}(r)=0.
	\end{equation}
\end{enumerate}

The equations satisfied by $\hat{\Psi}_\nu^X$ are formally identical to the equations above, with the modification that all physical quantities are replaced by their values corresponding to the false vacuum. 
For specific details on degeneracy, each explicit $\hat{\Psi}_\nu^X$ and the summation of determinants, please refer to the \textit{Details of Determinant Calculation} section of the \textit{Supplemental Material}. In the case of the Standard Model, if the Higgs potential is chosen as in Refs.~\cite{chigusaDecayRateElectroweak2018,Chigusa:2017dux}, the above formulas can also yield an analytical solution, and its limit results are consistent with those in Refs.~\cite{chigusaDecayRateElectroweak2018,Chigusa:2017dux,baratellaRevisingFullOneLoop2025}. In numerical calculations, we generally use the equation satisfied by $R_\nu^X(r)=\frac{\Psi^{X}_\nu(r)}{\hat{\Psi}_\nu^X(r)}$ to avoid divergence issues at infinity, allowing the results to converge more rapidly. The corresponding equations can be derived by substituting $\Psi^{X}_\nu(r)=R_\nu^X(r)\hat{\Psi}_\nu^X(r)$ into equations \eqref{equ-24} to \eqref{equ-27}. The results of equations and the Taylor expansion around the corresponding singularities are provided in the \textit{Numerical Computation Aspects} section of the \textit{Supplemental Material}.

Additionally, we may encounter zero mode issues in certain cases. 
For the Higgs field, we refrain from a detailed discussion of the dilatational zero modes here, as they occur only in the massless case, which can be handled following the approach described in Ref.~\cite{chigusaDecayRateElectroweak2018}. In the near‑scale‑invariant regime where the effective mass is very small, it can be approximated by the massless limit.  In the massive case, where such zero modes are absent, we simply replace the negative eigenvalue $R_1^\text{H}$ with $\bar{R}_1^{\text{H}}=\left|R_1^\text{H}\right| $. For $\nu=2$, the Higgs field always exhibits translational zero modes, which yield \cite{dunneFunctionalDeterminantsRadial2006,Ekstedt:2023sqc}
\begin{equation}
	\left(R_2^{\text{H}}\right)^{-\frac{\text{deg}_\text{H}(2;D)}{2}}=\frac{\mathcal{V}_{D}}{R^D}\left[\frac{2\left(2\pi\right)^{\frac{D}{2}-1}\phi_\infty \left|\phi_2\right|}{\kappa^2R^{2l-D+2}}\right]^{\frac{\text{deg}_\text{H}(2;D)}{2}},
\end{equation}
where a single zero mode of $R_2^{\text{H}}$ yields a prefactor $\frac{\mathcal{B}}{2\pi \kappa^2 R^{2l-D+2}}\int \mathrm{d}x$ originating from the path-integral measure over the collective coordinate, with $\frac{\mathcal{B}}{2\pi}$ being  canceled in the subsequent normalization, while the $D$ copies of $\int dx$ give $\frac{V_D}{R^D}$.
Thus, we can replace $R_2^\text{H}$ with
\begin{equation}
	\bar{R}_2^\text{H}\equiv \frac{\kappa^2R^{2l-D+2}}{2\left(2\pi\right)^{\frac{D}{2}-1} \phi_\infty \left|\phi_2\right|},
\end{equation}
and multiply $\frac{\mathcal{V}_{D}}{R^D}$ by $\mathcal{A}$.

For the gauge boson, $R^\text{Da}_1$ needs to be discussed separately. Following an approach similar to that in Ref.~\cite{endoFalseVacuumDecay2017}, for the case $v\neq0$, since the issue of symmetry breaking does not arise here, there is no zero-mode problem, and we can directly obtain $R^\text{Da}_1=\frac{v}{\phi_c}$ from the operator matrix.  For the case $v=0$, we obtain
\begin{equation}
	\left( R^\text{Da}_1\right)^{-\frac{n_G}{2}}=\mathcal{V}_G\left[\frac{\left(2\pi\right)^{\frac{D}{2}-1}\phi_c\phi_\infty}{\kappa^2R^{2l-D+2}}\right]^{\frac{n_G}{2}},
\end{equation}
where $n_G$ is the number of gauge bosons, and $\mathcal{V}_G$ is given by the product of the $n_G$ prefactors $\frac{1}{\kappa^2R^{2l-D+2}}\int \mathrm{d}\theta$ arising from the integration measures. Therefore, we can replace $R_1^\text{Da}$ with
\begin{equation}
	\bar{R}_1^\text{Da}\equiv \frac{\kappa^2R^{2l-D+2}}{\left(2\pi\right)^{\frac{D}{2}-1} \phi_c\phi_\infty},
\end{equation}
 and multiply $\mathcal{V}_{G}$ by $\mathcal{A}$.

\noindent\textit{ Renormalization and WKB Approximation--}
Now, we have unified $\ln \mathcal{A}_X$ into the form of \eqref{equ-16}. Unfortunately, such a summation is divergent and needs to be renormalized. We will present a universal method that theoretically allows for the calculation of arbitrary higher-order approximations in any dimension. 
This method is primarily based on the WKB approximation, the dimensional regularization and $\overline{\rm MS}$ scheme~\cite{Ekstedt:2023sqc,Ekstedt:2021kyx,Isidori:2001bm,Baacke:2003uw}, through which we truncate the approximation of the summation terms within the summation and then add back their known summation result.

On the one hand, we have previously expressed $R^X_\nu$ uniformly in the form of $\lim\limits_{r\rightarrow+\infty}\frac{\Psi^{X}_\nu(r)}{\hat{\Psi}^{X}_\nu(r)}$, where $\Psi^{X}_\nu(r)$ and $\hat{\Psi}^{X}_\nu(r)$ satisfy their individual equations. We can expand $\ln\Psi^{X}_\nu(r)$ (and $\ln\hat{\Psi}^{X}_\nu(r)$) in terms of  $\bar{\nu}\equiv\nu+\frac{D}{2}-2$ 
\begin{equation}\label{equ-38}
	\ln \Psi^{X}_\nu(r)=\sum_{n=0}^{+\infty}\int_{0}^{r}\d r C^X_n(r;D) \bar{\nu}^{1-n} ,
\end{equation}
where $C^X_n(r;D)$ can be obtained by substituting this expansion into the equations \eqref{equ-24} to \eqref{equ-27}.
On the other hand, the prefactor $\mathcal{A}_{X}$ can be written as
\begin{equation}\label{equ-37}
	\ln \mathcal{A}_{X}=\sum_{n=0}^{+\infty}\int_{0}^{r}\d r \text{Deg}_X(n;D)\delta C^X_n(r;D).
\end{equation}
Here, $\text{Deg}_X(n;D)\equiv\sum\limits_{\nu=1}^{+\infty}\text{deg}_X(\nu;D)\bar{\nu}^{1-n}$ is easily computable, and $\delta X\equiv X-\hat{X}$.

Since we have already expressed each formula in a dimension-dependent form, so we can directly replace $D$ in \eqref{equ-37} with $D-\epsilon$, and then simply multiply it by $\left( \frac{e^{\gamma_E}\mu^2R^2}{4\pi}\right) ^\frac{\epsilon}{2}$, then expand the result in $\epsilon$, and subtract $\frac{1}{\epsilon}r^{-\epsilon}\frac{\Omega_{D-\epsilon}}{\Omega_{D}}$ for each $\frac{1}{\epsilon}$ to obtain the corresponding add-back term. This procedure is ultimately equivalent to the result obtained by removing all the $\frac{1}{\epsilon}+\frac{1}{2}\ln 4\pi-\frac{1}{2}\gamma_E$ terms from expression
\begin{equation}\label{equ-39}
\begin{aligned}
	&\ln \mathcal{A}_{X}=\\
    &\sum_{n=0}^{+\infty}\int_{0}^{r}\d r \left( \mu Rr\right) ^\epsilon \frac{\Omega_D }{\Omega_{D-\epsilon}}\text{Deg}_X(n;D-\epsilon)\delta C^X_n(r;D-\epsilon) .
\end{aligned}
\end{equation}
Therefore, any part of \eqref{equ-38} containing the complete divergent term can be used as the subtraction term, while the corresponding part in \eqref{equ-39} can serve as the add-back term. 

Through concrete calculations, we find that the results of this method at $D=4$ match those in Ref.~\cite{baratellaFunctionalDeterminantsFalse2025} up to order $\bar{\nu}^{-2}$ and are more natural and intuitive than previous approaches~\cite{Isidori:2001bm,Chigusa:2017dux,chigusaDecayRateElectroweak2018,Shoji:2025jmg,baratellaRevisingFullOneLoop2025}. On the other hand, we have verified that the Higgs-sector finite parts also agree with the known scalar-bubble benchmarks of Ref.~\cite{dunneFunctionalDeterminantsRadial2006,k1b6-2gbs,Ekstedt:2023sqc} in any $D\geq3$, and the add-back finite parts have no $\mu$-dependence in odd D and are read off directly from the original expressions after pole subtraction, as correctly reflected in our results. Moreover, our method can be extended — after somewhat intricate but manageable calculations — to obtain results for arbitrary orders in any dimension. More details are given in the \textit{Details of WKB Approximation} sections of \textit{Supplemental Material}.
Additionally, we can find a method known as zeta function regularization in Ref.~\cite{dunneFunctionalDeterminantsRadial2006,k1b6-2gbs} whose equivalence with the dimensional regularization to some extent can be found in Refs.~\cite{Shoji:2025jmg,Bytsenko:2003tu}\footnote{
It should be noted here that instead of calculating the asymptotic expansion using the heat kernel expansion mentioned therein, we may also employ the WKB approximation as a substitute. However, since \eqref{equ-25} and \eqref{equ-26} are reduced equations, we cannot simply replace $\Delta_\nu$ with $\Delta_\nu-k^2$ for parameterization. Instead, we should use $\left( \mathcal{M}_X+k^2\right) \Psi_\nu^X=0$ and obtain the final position of $k$ in \eqref{equ-25} and \eqref{equ-26}.}. See also Ref.~\cite{Dunne:2005rt} for the renormalization method of angular momentum regularization and $\overline{\rm MS}$ scheme. 

\begin{figure}[!h]
	\centering
	\begin{minipage}{1\linewidth}
		\centering
		\includegraphics[width=0.8\linewidth]{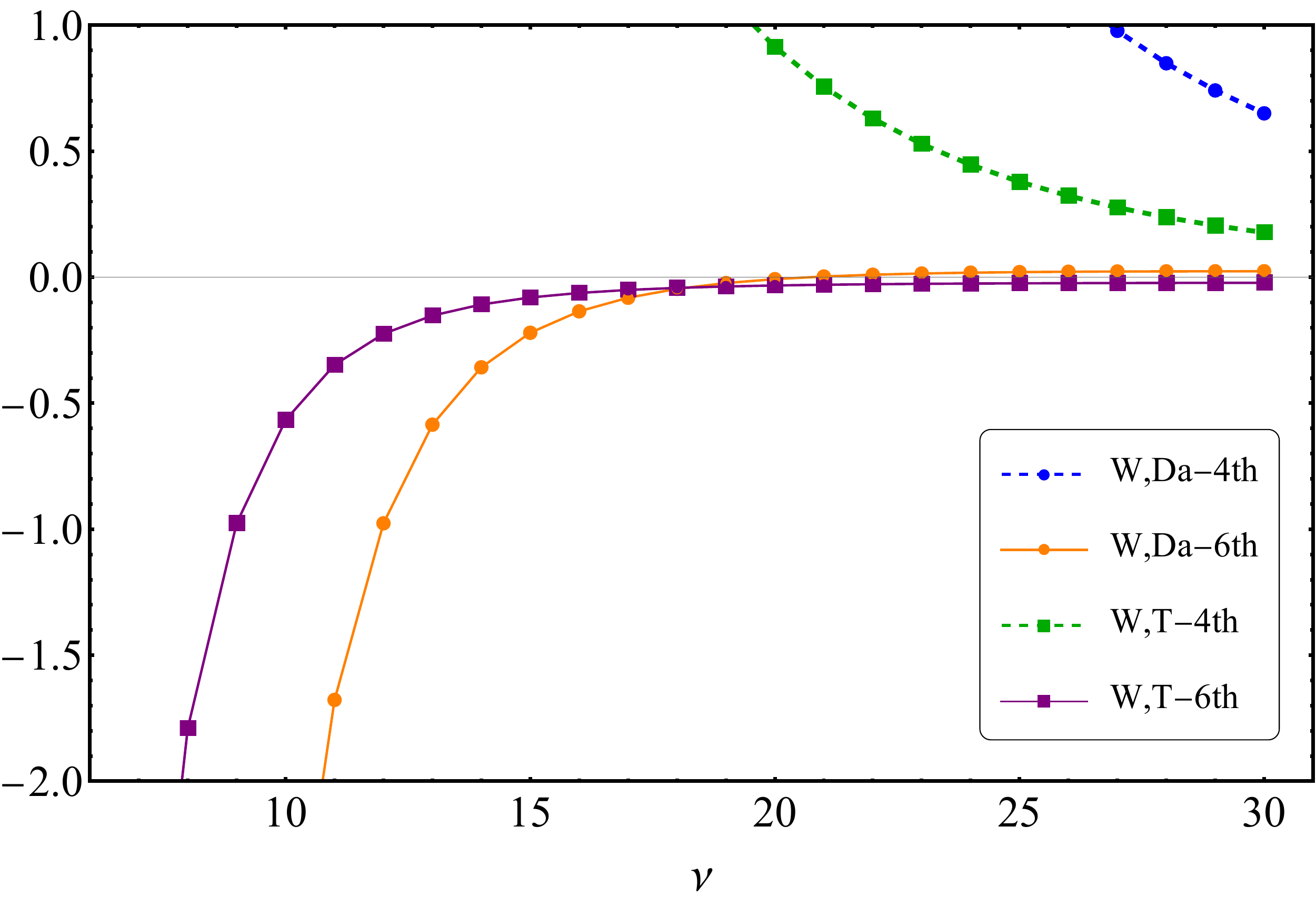}
	\end{minipage}
	\caption{Residuals of the WKB approximation for W boson($\times10^4$). X$-n$th represents the result obtained by subtracting its WKB approximation of X field up to order $n$. The result is not only scaled but also multiplied by $\bar{\nu}$ due to $\ln R^X_\nu\sim \delta C_1\bar{\nu}^{-1}+\dots$. And the factor by which the T mode part is multiplied is $\bar{\nu}+1$.}
	\label{fig-3}
\end{figure}

\noindent\textit{Numerical Examples--}
We provide two examples for the application of our method to the $D=4$ and $D=3$ theories. For $D=4$ theories, we consider the SMEFT with the potential being~\cite{Ekstedt:2023sqc}:
\begin{equation}
	V(\Phi)=\frac{1}{\kappa^2R^{4}}U(\bar{\phi})=\frac{\mu^4}{\lambda}\left(\frac{1}{2}\frac{m^2}{\mu^2}\bar{\phi}^2-\frac{1}{4} \bar{\phi}^4+\frac{1}{6}a \bar{\phi}^6\right),
\end{equation}
where $\kappa=\sqrt{\lambda}=0.378$, $R=\frac{1}{\mu}=1.097\times10^{-2}\text{GeV}^{-1}$ — however, the result is independent of the rescaling parameter $\kappa$ and $R$ — and $a=0.94$ . It can be easily seen that $v=0$ and $\hat{m}_\phi=\frac{m}{\mu}=0.241$. We can obtain the bounce solution via the shooting method, and find that $\phi_c=0.975$ and $\phi_\infty=47.962$. 
We choose our one-loop coupling constants as  $y_t=\frac{0.7317}{\sqrt{2}\kappa}=1.3687$, $g_W=\frac{0.6259}{2\kappa}=0.8279$ and $g_Z=\frac{0.7211}{2\kappa}=0.9539$ at the renormalization scale $\mu=m_Z$. As shown in Fig.~\ref{fig-3} , higher-order WKB approximations are more accurate, since the numerical deviations from lower orders become separable and can be removed. Finally, the vacuum decay rate in this case yields 
$\log_{10}(\gamma \times \text{Gyr} \times \text{Gpc}^3)=-282.9.$
Comparing with the Hubble constant at present $H_0^{-4}\simeq 10^3 \text{Gyr}\text{Gpc}^3$, it can be seen that the probability of transitioning from the electroweak symmetric vacuum to the broken one is negligible at zero temperature. Therefore,  the effect of temperature in the early universe should be taken into account for the false vacuum decay rate computation.  

We consider the three-dimensional action obtained from the four-dimensional spacetime using the dimensional reduction (DR) approach to the SMEFT\footnote{See Ref.~\cite{Ekstedt:2021kyx} for the study of solely scalar part contribution within the framework.} for the case of $D=3$ theory:
\begin{equation}
	\begin{aligned}
		S^{\text{light}}_{\text{3d}}=& \int {\d}^3 x \left[\frac{1}{4}F_{ij}^{a}F_{ij}^{a} + \left(D_{i}\Phi\right)^\dagger \left(D_{i}\Phi\right)+ V_{\text{3d}}\right],\\
		V_{\text{3d}} =&m_{3}^2\Phi^\dagger\Phi+\lambda_3(\Phi^\dagger\Phi)^2+c_{6,3}(\Phi^\dagger\Phi)^3,
	\end{aligned}
\end{equation}
where the covariant derivative is defined as $D_i=\partial_i-ig_3^aA_i^a\tau^a$. The relation between the 3D and 4D parameters can be found in Refs.~\cite{croonTheoreticalUncertaintiesCosmological2021,Chala:2025aiz}.
By setting the one-loop coupling constant and New physics scale
$\Lambda=570~\text{GeV}$, the reduced potential at $T=50~ \text{GeV}$ can be written as
\begin{equation}
	\begin{aligned}
		U_{\text{3d}}(\bar{\phi})=\frac{1}{2}\bar{\phi}^2-\frac{1}{4}\bar{\phi}^4+\frac{1}{6}a\bar{\phi}^6,
	\end{aligned}
\end{equation}
where $a=0.0848$. Note that here we use the four-dimensional re-scaling scheme with $R=0.0298~\text{GeV}^{-1}$ and $\kappa=2.70~\text{GeV}^{\frac{1}{2}}$. Similarly, from the numerical results, we can see that $\phi_{c}=3.102$, $\phi_{\infty}=10.87$, here
\begin{equation}
	\begin{aligned}
		\mathcal{B}=\frac{4\pi}{\kappa^2 R}\int_0^{+\infty} \d r r^2 \left[ \frac{1}{2}\left(\partial_{r}\bar{\phi}\right)^2+ U_{\text{3d}}(\bar{\phi})\right]=178.6\;.
	\end{aligned}
\end{equation}
Ignoring Langevin damping, we employ the following decay rate formula~\cite{Ekstedt:2023sqc}:
\begin{equation}
	\begin{aligned}
		\Gamma_T=\frac{\sqrt{\left|\lambda_{-}\right|}}{2\pi}\frac{\mathcal{V}_G}{R^4}\mathcal{A}_H^{-\frac{1}{2}}\mathcal{A}_W^{-1}\mathcal{A}_Z^{-\frac{1}{2}}e^{-\mathcal{B}},
	\end{aligned}
\end{equation}
where $\lambda_{-}$ is the rescaled negative eigenvalues of the Higgs' functional determinant. The final result can be obtained as $\ln(\Gamma_T\times\text{GeV} ^{-4})=-123.5$. Additionally, matching the 3D couplings  to the 4D case at the two-loop order~\cite{Liu:2026ask,Liu:2025ipj,Qin:2024dfp,Qin:2024idc} yields: $\ln(\Gamma_T\times\text{GeV} ^{-4})=-108.3$. Considering the fourth power of the Hubble constant during the radiation-dominated era $\ln(H^4\times\text{GeV} ^{-4})=4\ln(1.66\sqrt{g_*(T)}\frac{T^2}{M_{pl}}\times\text{GeV} ^{-1})\simeq-133.1$, the probability of transitioning from the false vacuum to the true vacuum is significant.  

\begin{figure}[!htp]
	\centering
	\begin{minipage}{1\linewidth}
		\centering
		\includegraphics[width=0.8\linewidth]{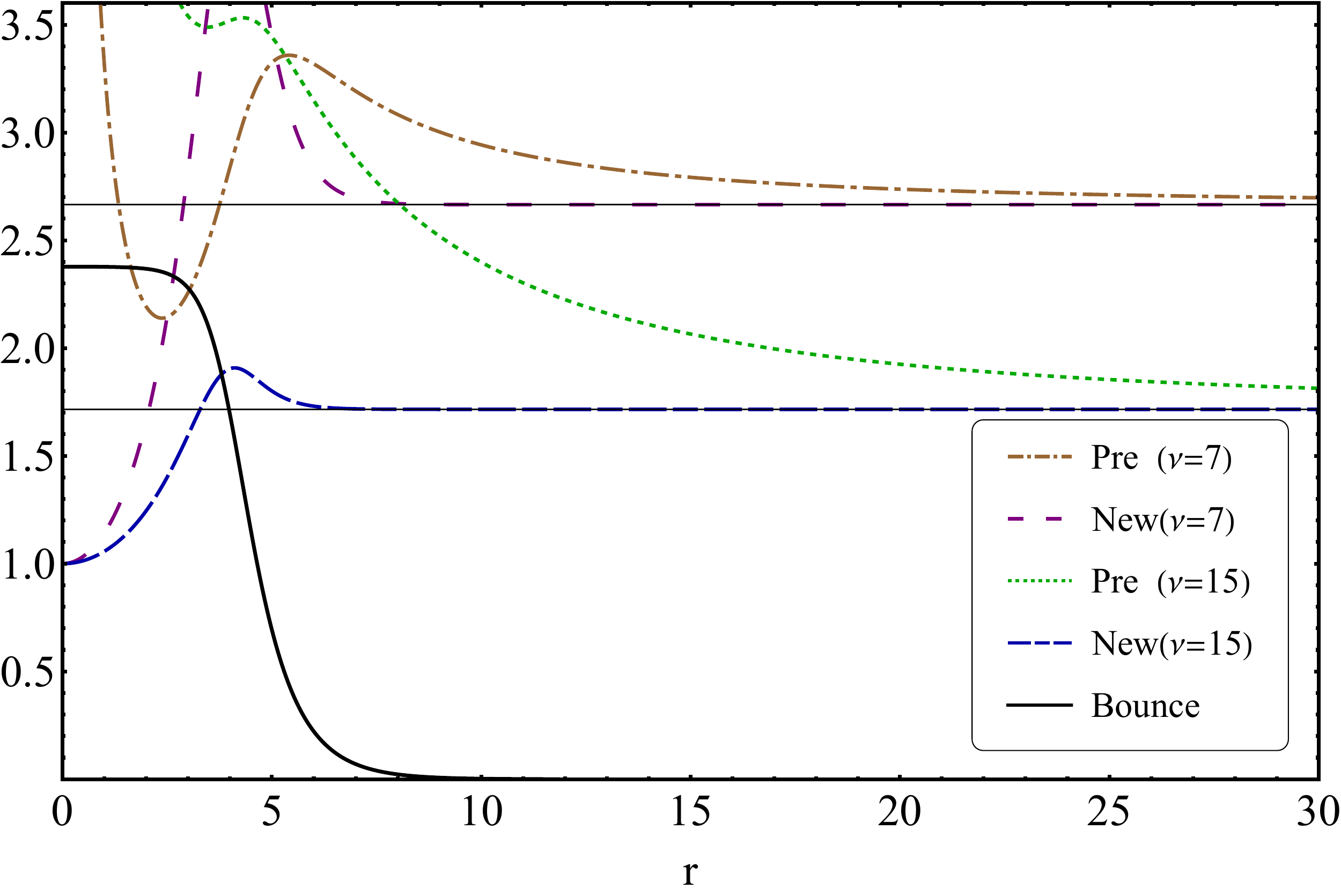}
	\end{minipage}
	\caption{Comparison of previous (denoted as ``Pre") and our new methods for the Da mode of W boson (tree level couplings). Here, the previous method employs the 3-dimensional version of the method described in Ref.~\cite{endoFalseVacuumDecay2017}.  }
	\label{fig-4}
\end{figure}

Fig.~\ref{fig-4} depicts that, in comparison with Ref.~\cite{endoFalseVacuumDecay2017}, our new method not only achieves good convergence but also avoids the need to perform three numerical calculations for each angular momentum to improve convergence.
Ref.~\cite{Ekstedt:2023sqc} takes the approximation that mixes Da mode and T mode by neglecting off-diagonal terms of the Da mode, while our framework preserves the correct mode structure and yields much more accurate gauge boson results.  Earlier attempts were conducted in the SU(2) Higgs model, where the heat-kernel expansion method fails in the thin-wall limit~\cite{Kripfganz:1994ha}, and Ref.~\cite{Baacke:1995bw} requires full matrix-element computation and divergent loop-diagram subtraction. 

\noindent\textit{ Conclusions--}
We present a general method for computing vacuum decay rates in D-dimensional electroweak theories. This extensible tool facilitates the study of metastable vacua, with applications to new physics beyond the Standard Model and cosmological phase transitions. By combining a WKB expansion with dimensional regularization, our approach allows systematic evaluation of high-order quantum corrections and remains numerically stable even for large angular momenta. The equations governing fluctuation determinants converge rapidly, offering a reliable framework for vacuum stability and decay analysis across dimensions. 

As a demonstration, we apply our method to the SMEFT in $D=4$ to incorporate new physics effects, and to its dimensionally reduced 
$D=3$ thermal counterpart to compute the bubble nucleation rate for a first-order EWPT. While obtaining the full result, including the dynamical prefactor of the decay rate, requires numerical simulations of the finite-temperature nucleation rate~\cite{Hirvonen:2024rfg,Hirvonen:2025hqn,Wang:2025ooq,Bian:2025twi}, our theoretical calculations pave the way toward a precise determination of the nucleation rate that underpins key predictions in early-universe cosmology: the gravitational wave signatures of first-order phase transitions~\cite{Mazumdar_2019,Caprini_2020,Athron_2024}, the origin of cosmic magnetic fields via magnetogenesis~\cite{PhysRevLett.51.1488,Di_2021,Enqvist:1993np,Yang_2022,Vachaspati:1991nm,Baym:1995fk,Grasso:1997nx}, and the viability of baryogenesis mechanisms accounting for the observed baryon asymmetry of the universe~\cite{Di:2024gsl,Di:2025ncl,Morrissey_2012,Cohen_2012}.

\noindent\textit{ Acknowledgements--}
This work is supported by the National Natural Science Foundation of China (NSFC) under Grants Nos.12322505, 12547101. We also acknowledge the Chongqing Natural Science Foundation under Grant No. CSTB2024NSCQ-JQX0022 and 
Chongqing Talents: Exceptional Young Talents Project No. cstc2024ycjh-bgzxm0020.

\bibliography{reference}

\end{document}


\onecolumngrid
\begin{center}
  \textbf{\large Supplemental Material}\\[.2cm]
\end{center}

In the supplementary materials, we present detailed design principles and computational specifics of the method. This part includes the calculation of the degeneracy, the equations of the reduced eigenfunction $\Psi^X_\nu$ and the specific expression of each $\hat{\Psi}_\nu^X$ (in \textit{Details of Determinant Calculation}), the calculations of coefficients for the WKB approximation (in \textit{Details of WKB Approximation}), the Taylor expansion of $\Psi_\nu^X$ near the singularity (at $r=0$) and the equations satisfied by $R_\nu^X$ (in \textit{Numerical Computation Aspects}).

\section{Details of Determinant Calculation}
The Higgs sector is straightforward. Thus, we will omit it here and only focus on the fermion and gauge boson sector. It should be noted that we have performed the following re-scalings for coordinates, fields, coupling constants, and the scale potential $U(\bar{\phi})$, respectively
\begin{equation}
	\begin{aligned}
		x\rightarrow Rx&\quad\left(y,g\right)\rightarrow \kappa\left(y,g\right),\\
		\left(\bar{\phi},h,\psi,A_\mu,a,c\right)&\rightarrow \frac{1}{\kappa R^l}\left(\bar{\phi},h,\frac{\psi}{R^{\frac{1}{2}}},A_\mu,a,c\right)\\
		U(\bar{\phi})&\rightarrow  \frac{1}{\kappa^2 R^{2l+2}}U(\bar{\phi}).
	\end{aligned}
\end{equation}

In addition, for  a more convenient unified expression, we choose the following auxiliary function 
\begin{equation}
	F_\nu(r;A)\equiv\left\lbrace \begin{matrix}
		\Gamma\left(\bar{\nu}+1\right) \left( \frac{2}{A}\right)^{\bar{\nu}} \frac{I_{\bar{\nu}}( A r)}{r^{\frac{D}{2}-1}},&\text{if}&A\neq0,\\
		r^{\nu-1},&\text{if}&A=0,
	\end{matrix} \right.
\end{equation}
where $I_k(x)$ is the modified Bessel function of the first kind.
\subsection{Fermion}
Due to the symmetry of the bounce solution, the contribution to the prefactor is given by
\begin{equation}
	\begin{aligned}
		\ln\mathcal{A}_{\psi}&=\ln\frac{\slashed{\partial}+\bar{\phi}_y}{\slashed{\partial}+\bar{v}_y}\\
		&=\frac{1}{2}\ln\frac{\left(\slashed{\partial}+\bar{\phi}_y\right)\left(-\slashed{\partial}+\bar{\phi}_y\right)}{\left(\slashed{\partial}+\bar{v}_y\right)\left(-\slashed{\partial}+\bar{v}_y\right)}\\
		&=\frac{1}{2}\ln\frac{-\partial^2+\gamma_r\bar{\phi}^\prime_y+\bar{\phi}_y^2}{-\partial^2+v_y^2},
	\end{aligned}
\end{equation}
where $\gamma_r=\frac{\slashed {x}}{r}$.

We need to discuss the cases where $D=2N$ and $D=2N+1$ separately. For the case where $D=2N$, expanding the fermion field using spherical harmonics can be viewed as the direct product of the spin representation $\left[\pm s\right]$ with an orbital component $\left[\nu-1\right]$. The result is a reducible representation
\begin{equation}
	\begin{aligned}
		\left[\nu-1\right]\otimes \left[+s\right]\simeq& \left[\nu-1,+s\right]\oplus\left[\nu-2,-s\right],\\
		\left[\nu-1\right]\otimes \left[-s\right]\simeq& \left[\nu-1,-s\right]\oplus\left[\nu-2,+s\right],
	\end{aligned}
\end{equation}
where the highest weight of $\left[\nu-1,+s\right]$ and $\left[\nu-1,-s\right]$ are $(\nu-1+\frac12, \frac12, \ldots, \frac12)$ and $(\nu-1+\frac12, \frac12, \ldots, -\frac12)$, respectively. Therefore, the eigenfunctions of fermion field can be expanded in an orthogonal basis as follows:
\begin{equation}
	\begin{aligned}
		\Psi^\psi(r)=\sum_{\nu=1}^{+\infty}\left(a_{1,\nu}(r)\xi^+_{1,\nu}+a_{2,\nu}(r)\xi^-_{2,\nu+1}\right)+(+\leftrightarrow-,a\rightarrow b),
	\end{aligned}
\end{equation}
where $\gamma_r\xi^{\pm}_{1,\nu}=\xi^{\mp}_{2,\nu+1}$. We obtain
\begin{equation}
	\begin{aligned}
		\begin{pmatrix}
			-\Delta_\nu+\bar{\phi}^2_y&\bar{\phi}_y^\prime\\
			\bar{\phi}_y^\prime&-\Delta_{\nu+1}+\bar{\phi}^2_y
		\end{pmatrix}
		\begin{pmatrix}
			a_{1,\nu}(r)\\
			a_{2,\nu}(r)
		\end{pmatrix}&=0,\\
		\begin{pmatrix}
			-\Delta_\nu+\bar{\phi}^2_y&\bar{\phi}_y^\prime\\
			\bar{\phi}_y^\prime&-\Delta_{\nu+1}+\bar{\phi}^2_y
		\end{pmatrix}
		\begin{pmatrix}
			b_{1,\nu}(r)\\
			b_{2,\nu}(r)
		\end{pmatrix}&=0.
	\end{aligned}
\end{equation}

The corresponding degeneracy can be given by calculating the dimension of the representation with the highest weight $(\nu-1+\frac12, \frac12, \ldots, \pm\frac12)$ in the Lie algebra of type $D_N$ using the Weyl dimension formula. Since the degeneracies for $(\nu-1+\frac12, \frac12, \ldots, \frac12)$ and $(\nu-1+\frac12, \frac12, \ldots, -\frac12)$ are identical, and the equations governing the corresponding $a_{i,\nu}(r)$ and $b_{i,\nu}(r)$ are identical in form, the complete result can be obtained by simply multiplying the result for $(\nu-1+\frac12, \frac12, \ldots, \frac12)$ by 2. We obtain
\begin{equation}
	\text{deg}_\psi(\nu;2N)=2^{N}\frac{\Gamma(\nu+D-2)}{\Gamma(\nu)\Gamma(D-1)}.
\end{equation}

For the case where $D=2N+1$, the reducible representation can be written as
\begin{equation}
	\begin{aligned}
		\left[\nu-1\right]\otimes \left[s\right]\simeq [\nu-1,s]\oplus[\nu-2,s].
	\end{aligned}
\end{equation}

Similarly, the eigenfunctions of fermion field can be expanded in an orthogonal basis as
\begin{equation}
	\begin{aligned}
		\Psi^\psi(r)=\sum_{\nu=1}^{+\infty}\left(a_{1,\nu}(r)\xi_{1,\nu}+a_{2,\nu}(r)\xi_{2,\nu+1}\right),
	\end{aligned}
\end{equation}
where $\gamma_r\xi_{1,\nu}=\xi_{2,\nu+1}$. We obtain
\begin{equation}
	\begin{aligned}
		\begin{pmatrix}
			-\Delta_\nu+\bar{\phi}^2_y&\bar{\phi}_y^\prime\\
			\bar{\phi}_y^\prime&-\Delta_{\nu+1}+\bar{\phi}^2_y
		\end{pmatrix}
		\begin{pmatrix}
			a_{1,\nu}(r)\\
			a_{2,\nu}(r)
		\end{pmatrix}&=0.
	\end{aligned}
\end{equation}

The corresponding degeneracy can be given by the dimension of the representation with the highest weight $(\nu-1+\frac12, \frac12, \ldots, \frac12)$ in the Lie algebra of type $B_N$
\begin{equation}
	\begin{aligned}
		\text{deg}_\psi(\nu;2N+1)=2^{N}\frac{\Gamma(\nu+D-2)}{\Gamma(\nu)\Gamma(D-1)}.
	\end{aligned}
\end{equation}

Therefore, we obtain
\begin{equation}
	\begin{aligned}
		\ln\mathcal{A}_\psi=\sum_{\nu=1}^{+\infty}\frac{1}{2}\text{deg}_\psi(\nu;D)\left[\ln\left(\lim_{r\rightarrow+\infty}\frac{\det\left[\Psi^\psi_{\nu,1},\Psi^\psi_{\nu,2}\right]}{\det\left[\hat{\Psi}^\psi_{\nu,1},\hat{\Psi}^\psi_{\nu,2}\right]}\right)-\ln\left(\lim_{r\rightarrow0}\frac{\det\left[\Psi^\psi_{\nu,1},\Psi^\psi_{\nu,2}\right]}{\det\left[\hat{\Psi}^\psi_{\nu,1},\hat{\Psi}^\psi_{\nu,2}\right]}\right)\right],
	\end{aligned}
\end{equation}
where
\begin{equation}
	\begin{aligned}
		&\text{deg}_\psi(\nu;D)=2^{\lfloor\frac{D}{2}\rfloor}\frac{\Gamma(\nu+D-2)}{\Gamma(\nu)\Gamma(D-1)},\\
		&\begin{pmatrix}
			-\Delta_\nu+\bar{\phi}^2_y&\bar{\phi}_y^\prime\\
			\bar{\phi}_y^\prime&-\Delta_{\nu+1}+\bar{\phi}^2_y
		\end{pmatrix}\Psi^\psi_{\nu,i}=0,\\
		&\begin{pmatrix}
			-\Delta_{\nu}+v^2_y&0\\
			0& -\Delta_{\nu+1}+v^2_y
		\end{pmatrix}\hat{\Psi}^\psi_{\nu,i}=0.
	\end{aligned}
\end{equation}

The solution for $\hat{\Psi}^\psi_{\nu,i}$ can also be readily obtained as follows
\begin{equation}
	\det\left[\hat{\Psi}^\psi_{\nu,1},\hat{\Psi}^\psi_{\nu,2} \right] =\left\lbrace\begin{matrix}
		\Gamma\left(\nu+\frac{D}{2}-1\right) \Gamma\left(\nu+\frac{D}{2}\right) \left( \frac{2}{v_t}\right)^{2\nu+D-3} \frac{I_{\nu+\frac{D}{2}-2}( v_t r)I_{\nu+\frac{D}{2}-1}( v_t r)}{r^{D-2}},&\text{if}&v_t\neq0,\\
		r^{2\nu-1},&\text{if}&v_t=0.
	\end{matrix} \right. 
\end{equation}

On the other hand, the formal solution for $\Psi^T_{\nu,i}$ can be written in form as
\begin{equation}
	\det\left[\Psi^\psi_{\nu,1},\Psi^\psi_{\nu,2}\right]=\left| \begin{matrix}
		\frac{1}{\bar{\phi}_y}\frac{1}{r^{\nu+D-2}}\partial_rr^{\nu+D-2}f&\frac{1}{\bar{\phi}_y}\frac{1}{r^{\nu+D-2}}\partial_rr^{\nu+D-2}g-\frac{1}{\bar{\phi}_y^2}\frac{1}{r^{\nu+D-2}}\partial_rr^{\nu+D-2}f\\
		f&g
	\end{matrix}\right|,
\end{equation}
with
\begin{equation}\label{equ-2.27}
	\begin{aligned}
		\left(-\Delta_{\nu+1}+\frac{\bar{\phi}_y^\prime}{\bar{\phi}_y} \frac{1}{r^{\nu+D-2}}\partial_rr^{\nu+D-2}+\bar{\phi}_y^2\right)f&=0,\\
		\left(-\Delta_{\nu+1}+\frac{\bar{\phi}_y^\prime}{\bar{\phi}_y} \frac{1}{r^{\nu+D-2}}\partial_rr^{\nu+D-2}+\bar{\phi}_y^2\right) g&=\frac{\bar{\phi}_y^\prime}{\bar{\phi}_y^2}\frac{1}{r^{\nu+D-2}}\partial_rr^{\nu+D-2}f.
	\end{aligned}
\end{equation}

However, this determinant depends solely on $f$. Actually, we can simply convert it into the integral form
\begin{equation}\label{equ-2.28}
	\begin{aligned}
		\det\left[\Psi^\psi_{\nu,1},\Psi^\psi_{\nu,2}\right]&=\frac{1}{\bar{\phi}_yr^{\nu+D-2}}\left[\frac{f\partial_r\left( r^{\nu+D-2}f\right) }{\bar{\phi}_y}-r^{\nu+D-2}\left(f\partial_r g-g\partial_rf\right)\right]\\
		&=\frac{f\partial_r\left( r^{\nu+D-2}f\right)}{\bar{\phi}_y^2r^{\nu+D-2}}+\frac{1}{r^{D-1}}\int_{0}^{r}\frac{\bar{\phi}_y^\prime}{\bar{\phi}_y^3}\frac{f\partial_r\left( r_1^{\nu+D-2}f\right)}{r_1^{\nu-1}}\d r_1\\
		&=\lim_{r\rightarrow0}\frac{f\partial_r\left( r^{\nu+D-2}f\right)}{\bar{\phi}_y^2r^{\nu+D-2}}+\frac{1}{r^{D-1}}\int_{0}^{r}\frac{1}{\bar{\phi}_y}\partial_r\left[\frac{f\partial_r\left( r_1^{\nu+D-2}f\right)}{\bar{\phi}_yr_1^{\nu-1}}\right]\d r_1.\\
	\end{aligned}
\end{equation}

Due to the limit condition at $r\rightarrow 0$, $\left. f\right|_{r\rightarrow0} \sim r^\nu$ and the previous limit in \eqref{equ-2.28} thus becomes 0. We obtain
\begin{equation}
	\begin{aligned}
		\lim_{r\rightarrow0}\frac{\det\left[\Psi^\psi_{\nu,1},\Psi^\psi_{\nu,2}\right]}{\det\left[\hat{\Psi}^\psi_{\nu,1},\hat{\Psi}^\psi_{\nu,2}\right]}&=\lim_{r\rightarrow0}\frac{1}{r^{2\nu-1}}\left[\frac{f\partial_r\left( r^{\nu+D-2}f\right)}{\bar{\phi}_y^2r^{\nu+D-2}}+\frac{1}{r^{D-1}}\int_{0}^{r}\frac{\bar{\phi}_y^\prime}{\bar{\phi}_y^3}\frac{f\partial_r\left( r_1^{\nu+D-2}f\right)}{r_1^{\nu-1}}\d r_1\right]=\frac{2\nu+D-2}{\phi_{cy}^2}.
	\end{aligned}
\end{equation}

The asymptotic behavior at $r\rightarrow+\infty$ can be obtained using L'Hôpital's rule. When $v\neq0$, we have
\begin{equation}
	\begin{aligned}
		&\Delta_{\nu+1}f\sim v_y^2f\quad\Rightarrow\quad \left. f\right|_{r\rightarrow+\infty} \sim K\frac{I_{\nu+\frac{D}{2}-1}\left( v_yr\right) }{r^{\frac{D}{2}-1}},\\
		\lim_{r\rightarrow+\infty}\frac{\det\left[\Psi^\psi_{\nu,1},\Psi^\psi_{\nu,2}\right]}{\det\left[\hat{\Psi}^\psi_{\nu,1},\hat{\Psi}^\psi_{\nu,2}\right]}
		&=\frac{1}{2\nu+D-2}\lim_{r\rightarrow+\infty}\frac{v_y}{ F_\nu(r;v_y)^2}\frac{1}{r^{D-1}}\int_{0}^{r}\frac{1}{\bar{\phi}_y}\partial_r\left[\frac{f\partial_r\left( r_1^{\nu+D-2}f\right)}{\bar{\phi}_yr_1^{\nu-1}}\right]\d r_1\\
		&=\frac{1}{2\nu+D-2}\lim_{r\rightarrow+\infty}\frac{1}{2F_\nu(r;v_y)^2}\left\lbrace f^2+\left[\frac{ \partial_r\left( r^{\nu+D-2}f\right) }{\bar{\phi}_yr^{\nu+D-2}}\right]^2\right\rbrace\\
		&=\frac{1}{2\nu+D-2}\left[\lim_{r\rightarrow+\infty}\frac{\partial_r\left( r^{\nu+D-2}f\right) }{\bar{\phi}_yF_\nu(r;v_y)r^{\nu+D-2}}\right]^2.
	\end{aligned}
\end{equation}

When $v=0$, we have
\begin{equation}
	\begin{aligned}
		\partial_r\left[\frac{\partial_r\left(r^{\nu+D-2}f\right) }{r^{2\nu+D-3}\bar{\phi}_y} \right]&=\frac{\bar{\phi}_yf}{r^{\nu-1}}\Rightarrow \frac{\partial_r\left(r^{\nu+D-2}f\right) }{fr^{\nu+D-2}\bar{\phi}_y}\sim \frac{\bar{\phi}_yfr^{-\nu+1}}{\partial_r\left(fr^{-\nu+1}\right)},\\
		\lim_{r\rightarrow+\infty}\frac{\det\left[\Psi^\psi_{\nu,1},\Psi^\psi_{\nu,2}\right]}{\det\left[\hat{\Psi}^\psi_{\nu,1},\hat{\Psi}^\psi_{\nu,2}\right]}
		&=\lim_{r\rightarrow+\infty}\frac{1}{F_\nu(r;v_y)^2}\frac{1}{r^{D}}\int_{0}^{r}\frac{1}{\bar{\phi}_y}\partial_r\left[\frac{f\partial_r\left( r_1^{\nu+D-2}f\right)}{\bar{\phi}_yr_1^{\nu-1}}\right]\d r_1\\
		&=\frac{1}{2\nu+D-2}\left[\lim_{r\rightarrow+\infty}\frac{\partial_r\left( r^{\nu+D-2}f\right) }{\bar{\phi}_yF_\nu(r;v_y)r^{\nu+D-2}}\right]^2.
	\end{aligned}
\end{equation}

Therefore, setting $\Psi^\psi_\nu(r)=\frac{\phi_{c}\partial_r\left( r^{\nu+D-2}f\right) }{\left(2\nu+D-2\right)\bar{\phi}r^{\nu+D-2}}$, the first equation in \eqref{equ-2.27} becomes
\begin{equation}
	\begin{aligned}
		\left( \Delta_\nu-\bar{\phi}^2_y\right) \Psi^\psi_\nu(r)&=\frac{\bar{\phi}_y^\prime}{\bar{\phi}_y}\left(\partial_r-\frac{\nu-1}{r}\right)\Psi^\psi_\nu(r),\\
	\end{aligned}
\end{equation}
and
\begin{equation}
	\begin{aligned}
		\ln\mathcal{A}_{\psi}&=\sum_{\nu=1}^{\infty}\text{deg}_\psi(\nu;D) \ln\left( \frac{\Psi^\psi_\nu}{\hat{\Psi}^\psi_\nu}\right).
	\end{aligned}
\end{equation}

\subsection{Gauge boson}
The result for the gauge boson can be derived from the following decomposition of $A_\mu$
\begin{equation}
	\begin{aligned}
		A_\mu=\sum_{n=1}^{+\infty}\left[a_{1,n}(r)\hat{x}_\mu+\frac{a_{2,n}(r)}{L}r\partial_\mu+\sum_{i=3}^{D}a_{i,n}(r)\epsilon_{\mu\nu\rho\sigma}V_{\nu}^{i}x_\rho\partial_\sigma\right]Y_n(\theta).
	\end{aligned}
\end{equation}

The contribution of gauge boson to the prefactor can be written as
\begin{equation}
	\ln\mathcal{A}_{A,a}=\sum_{\nu=1}^{\infty}\left[\text{deg}_{\text{Da}}(\nu;D)\ln \frac{\mathcal{M}_\nu^{\text{Da}}}{\widehat{\mathcal{M}}_\nu^{\text{Da}}}+\text{deg}_{\text{T}}(\nu;D)\ln \frac{\mathcal{M}_\nu^{\text{T}}}{\widehat{\mathcal{M}}_\nu^{\text{T}}} \right],
\end{equation}
where
\begin{equation}
	\begin{aligned}
		\mathcal{M}_{\nu=1}^{\text{Da}}=&\begin{pmatrix}
			-\frac{1}{\xi}\Delta_2+ \bar{\phi}^2_g&\bar{\phi}_g^\prime-\bar{\phi}_g\partial_r\\
			2\bar{\phi}_g^{\prime}+\bar{\phi}_g\frac{1}{r^{D-1}}\partial_rr^{D-1}&-\Delta_1+\frac{\Delta_1\bar{\phi}_g}{\bar{\phi}_g}
		\end{pmatrix},\\
		\mathcal{M}_{\nu>1}^{\text{Da}}=&\begin{pmatrix}
			-\Delta_\nu+\frac{D-1}{r^2}+\bar{\phi}_g^2&-\frac{2L}{r^2}&\bar{\phi}_g^{\prime}-\bar{\phi}_g\partial_r\\
			-\frac{2L}{r^2}&-\Delta_\nu-\frac{D-3}{r^2}+\bar{\phi}^2_g&-\frac{L}{r}\bar{\phi}_g\\
			2\bar{\phi}_g^{\prime}+\bar{\phi}_g\frac{1}{r^{D-1}}\partial_rr^{D-1}&-\frac{L}{r}\bar{\phi}_g&-\Delta_\nu+\frac{\Delta_1\bar{\phi}_g}{\bar{\phi}_g}\\
		\end{pmatrix}+\left(1-\frac{1}{\xi}\right)\begin{pmatrix}
			\Delta_2&-L\partial_r\frac{1}{r}&0\\
			\frac{L}{r^D}\partial_rr^{D-1}&-\frac{L^2}{r^2}&0\\
			0&0&0
		\end{pmatrix},
		\\
		\mathcal{M}_{\nu}^{\text{T}}=&-\Delta_{\nu+1}+\bar{\phi}^2_g.
	\end{aligned}
\end{equation}

Since the Da modes correspond one-to-one with scalar bosons, its degeneracy is the same as that of the spherical harmonics. For $\text{deg}_\text{T}(\nu;D)$, note that the direct product of the vector field and the angular momentum component can be decomposed as
\begin{equation}
	\left[\nu-1\right]\otimes \mathcal{V}\simeq \left[\nu\right]\oplus\left[\nu-2\right]\oplus \mathcal{T},
\end{equation}
where $\mathcal{T}$ is the transverse part, whose highest weight is $(\nu-1,1,0,0,\ldots)$. Thus, the degeneracy of this part satisfies (Note that we have shifted the labeling of the transverse components.)
\begin{equation}
		\text{deg}_\text{T}(\nu;D)=D\text{deg}_\text{Da}(\nu+1;D)-\text{deg}_\text{Da}(\nu+2;D)-\text{deg}_\text{Da}(\nu;D).
\end{equation}

The T mode sector, being analogous to the Higgs sector, will also be omitted here. Moreover, since the case of $\mathcal{M}_{\nu=1}^{Da}$ closely resembles the four-dimensional discussion in Ref.~\cite{endoFalseVacuumDecay2017}, and its conclusion has already been presented in the main text, we will focus exclusively on the case of $\mathcal{M}_{\nu>1}^{Da}$ here.

For the case of $\mathcal{M}_{\nu>1}^{Da}$, the solution can be formally written as
\begin{equation}
	\det\Psi_\nu=\left|\begin{matrix}
		\left(\nu-1\right)r^{\nu-2}&\partial_r\chi_2+\frac{L}{r}\frac{1}{\bar{\phi}_g^2}\eta_2&\partial_r\chi_3+\frac{L}{r}\frac{1}{\bar{\phi}_g^2}\eta_3-\frac{2\bar{\phi}_g^\prime}{\bar{\phi}_g^3}\zeta_3\\
		Lr^{\nu-2}&\frac{L}{r}\chi_2+\frac{1}{\bar{\phi}_g^2}\frac{1}{r^{D-2}}\partial_rr^{D-2}\eta_2&\frac{L}{r}\chi_3+\frac{1}{\bar{\phi}_g^2}\frac{1}{r^{D-2}}\partial_rr^{D-2}\eta_3\\
		\bar{\phi}_gr^{\nu-1}&\bar{\phi}_g\chi_2&\bar{\phi}_g\chi_3+\frac{1}{\bar{\phi}_g}\zeta_3
	\end{matrix}\right|,
\end{equation}
with ($i=2,3$)
\begin{equation}\label{equ-76}
	\begin{aligned}
		\Delta_\nu\chi_i&=\frac{2L}{r}\frac{\bar{\phi}_g^\prime}{\bar{\phi}_g^3}\eta_i+\frac{1}{r^{D-1}}\partial_r\frac{2\bar{\phi}_g^\prime}{\bar{\phi}_g^3}\zeta_ir^{D-1}-\xi\zeta_i,\\
		\left(\Delta_\nu-2\frac{\bar{\phi}_g^\prime}{\bar{\phi}_g}\frac{1}{r^{D-2}}\partial_{r}r^{D-2}-\bar{\phi}_g^2\right)\eta_i&=-\frac{2L}{r}\frac{\bar{\phi}_g^\prime}{\bar{\phi}_g}\zeta_i,\\
		\zeta_i&={0,r^{\nu-1}}.
	\end{aligned}
\end{equation}

Ref.~\cite{endoFalseVacuumDecay2017} discussed that this result depends solely on $\eta_2$ in $D=4$, and we can provide another proof. Actually, $\det\Psi_\nu$ can be written as
\begin{equation}
	\det\Psi_\nu
	=\frac{r^{\nu+1-D}}{\bar{\phi}_g} \left|\begin{matrix}
		r^{\nu-1}\partial_r\left( \frac{\chi_2}{r^{\nu-1}}\right) +\frac{L}{r}\frac{\eta_2}{\bar{\phi}^2_g} &r^{\nu-1}\partial_r\left( \frac{\chi_3}{r^{\nu-1}}\right)+\frac{L}{r}\frac{\eta_3}{\bar{\phi}^2_g}+\left( \partial_r\frac{1}{\bar{\phi}^2_g}\right) \zeta_3-\frac{1}{\bar{\phi}^2_g}\partial_r\zeta_3\\
		\partial_r\left( \eta_2r^{D-2}\right)&\partial_r\left( \eta_3r^{D-2}\right)-L\zeta_3r^{D-3}
	\end{matrix}\right|\equiv\frac{r^{\nu+1-D}}{\bar{\phi}_g}\left|\begin{matrix}
		P_2&P_3\\
		Q_2&Q_3
	\end{matrix}\right|.
\end{equation}

To simplify this determinant, by using Equation \eqref{equ-76}, we note that
\begin{equation}
	\begin{aligned}
		\partial_r\left[ r^{\nu+D-2}P_i\right] 
		&=\frac{L}{\bar{\phi}^2_g}\partial_r\left(\eta_ir^{\nu+D-3}\right)-\left(2\nu+D-4\right)\left(\nu-1\right)\frac{1}{\bar{\phi}^2_g}\zeta_i r^{\nu+D-4}-\xi\zeta_i r^{\nu+D-2},\\
		\partial_r\left[\frac{L}{\nu+D-3}\frac{r^{\nu}}{\bar{\phi}^2_g} Q_i\right]&=\frac{L}{\bar{\phi}^2_g} \partial_r\left(\eta_ir^{\nu+D-3}\right)
		+\frac{L\eta_ir^{\nu+D-2}}{\nu+D-3}
		-\left(2\nu+D-4\right)\left(\nu-1\right)\frac{1}{\bar{\phi}^2_g}\zeta_ir^{\nu+D-4}.
	\end{aligned}
\end{equation}

We obtain
\begin{equation}
	P_i-\frac{L}{\nu+D-3}\frac{1}{r^{D-2}\bar{\phi}^2_g}Q_i=-\frac{1}{r^{\nu+D-2}}\frac{L}{\nu+D-3}\int_{0}^{r}\eta_ir_1^{\nu+D-2}\d r_1-\frac{\xi\zeta_ir}{2\nu+D-2}.
\end{equation}

Therefore,
\begin{equation}
	\begin{aligned}
		\det\Psi_\nu=&\frac{1}{\left(\nu+D-3\right) r^{2D-3}\bar{\phi}_g}\left|\begin{matrix}
			-L\int_{0}^{r}\eta_2r_1^{\nu+D-2}\d r_1&-L\int_{0}^{r}\eta_3r_1^{\nu+D-2}\d r_1-\frac{\nu+D-3}{2\nu+D-2}\xi r^{2\nu+D-2}\\
			\partial_r\left( \eta_2r^{D-2}\right)&\partial_r\left( \eta_3r^{D-2}\right)-Lr^{\nu+D-4}
		\end{matrix}\right|\\
		=&\frac{1}{\left(\nu+D-3\right) r^{2D-3}\bar{\phi}_g}\left(\left|\begin{matrix}
			L\int_{0}^{r}\eta_2r_1^{\nu+D-2}\d r_1&-\frac{\nu+D-3}{2\nu+D-2}\xi r^{2\nu+D-2}\\
			\partial_r\left( \eta_2r^{D-2}\right)&Lr^{\nu+D-4}
		\end{matrix}\right|\right. \\
		&\left. -L\left|\begin{matrix}
			\int_{0}^{r}\eta_2r_1^{\nu+D-2}\d r_1&\int_{0}^{r}\eta_3r_1^{\nu+D-2}\d r_1\\
			\partial_r\left( \eta_2r^{D-2}\right)&\partial_r\left( \eta_3r^{D-2}\right)
		\end{matrix}\right|\right) .
	\end{aligned}
\end{equation}

On the other hand, we have
\begin{equation}
	\eta_3r^{D-2}=\eta_2r^{D-2}\int_{0}^{r}\frac{L\bar{\phi}^2_g\d r_1}{\eta_2^2r^{D-1}_1} \int_{0}^{r_1}\eta_2\left( \partial_r\frac{1}{\bar{\phi}^2_g}\right)r_2^{\nu+D-3}\d r_2.
\end{equation}

Thus,
\begin{equation}
	\begin{aligned}
		&\left|\begin{matrix}
			\int_{0}^{r}\eta_2r_1^{\nu+D-2}\d r_1&\int_{0}^{r}\eta_3r_1^{\nu+D-2}\d r_1\\
			\partial_r\left( \eta_2r^{D-2}\right)&\partial_r\left( \eta_3r^{D-2}\right)
		\end{matrix}\right|\\
		&=\partial_r\left(\eta_3r^{D-2}\int_{0}^{r}\eta_2r_1^{\nu+D-2} \d r_1-\eta_2r^{D-2}\int_{0}^{r}\eta_3r_1^{\nu+D-2} \d r_1 \right)\\
		&=L\partial_r\left[ \eta_2r^{D-2}\int_{0}^{r}\frac{\bar{\phi}^2_g\d r_1}{\eta_2^2r_1^{D-1}} \left( \int_{0}^{r_1}\eta_2\left( \partial_r\frac{1}{\bar{\phi}^2_g}\right)r_2^{\nu+D-3}\d r_2\right)\left( \int_{0}^{r_1}\eta_2r_3^{\nu+D-2} \d r_3\right)  \right],\\
	\end{aligned}
\end{equation}
and
\begin{equation}
	\begin{aligned}
		\det\Psi_\nu
		&=\frac{r^{2\nu-D+1}\partial_r\left( \eta_2r^{D-2}\right)}{\bar{\phi}_g} \left\lbrace \frac{\left(\nu-1\right)\bar{\phi}^2_g}{\eta_2 r^{2\nu+D-1}\partial_r\left( \eta_2r^{D-2}\right)} \left[ \int_{0}^{r}\frac{\partial_r\left(\eta_2r_1^{\nu+D-3}\right)}{\bar{\phi}^2_g} \d r_1\right]\left( \int_{0}^{r}\eta_2r_2^{\nu+D-2} \d r_2\right)\right. \\
		&\left. +\frac{2\left(\nu-1\right)}{ r^{2\nu+D-2}} \int_{0}^{r}\frac{\bar{\phi}^2_g\d r_1}{\eta^2_2r_1^{D-1}} \left( \int_{0}^{r_1}\frac{\eta_2r_2^{\nu+D-3}\bar{\phi}^\prime_g}{\bar{\phi}^3_g}\d r_2\right)\left( \int_{0}^{r_1}\eta_2r_3^{\nu+D-2} \d r_3\right)+\frac{\xi}{2\nu+D-2}\right\rbrace\\
		&\equiv\frac{r^{2\nu-D+1}\partial_r\left( \eta_2r^{D-2}\right)}{\left(2\nu+D-2\right)\bar{\phi}_g}\left( I_1 + I_2 +\xi  \right).
	\end{aligned}
\end{equation}

We can obtain asymptotic behavior of $I_1 + I_2$ in $r\rightarrow0$ and $r\rightarrow+\infty$ using L'Hôpital's rule
\begin{equation}
	\begin{aligned}
		I_2&\sim \frac{2\left( \nu-1\right)\bar{\phi}^2_g }{ \eta^2_2r^{2\nu+2D-4}}\left( \int_{0}^{r}\frac{\eta_2r_1^{\nu+D-3}\bar{\phi}^\prime_g}{\bar{\phi}^3_g}\d r_1\right)\left( \int_{0}^{r}\eta_2r_3^{\nu+D-2} \d r_3\right),\\
		I_1 + I_2&\sim \frac{\left( \nu-1\right)\int_{0}^{r}\eta_2r_1^{\nu+D-2} \d r_1}{\eta_2r^{\nu+D-1}}\left[\frac{\partial_r\left( \eta_2r^{2\nu+2D-4}\right)}{r^{2\nu+D-2}\partial_r\left( \eta_2r^{D-2}\right)}\frac{\bar{\phi}^2_g }{\eta_2r^{\nu+D-3}}\int_{0}^{r}\frac{\partial_r\left(\eta_2r_1^{\nu+D-3}\right)}{\bar{\phi}^2_g}\d r_1-1\right].
	\end{aligned}
\end{equation}

We can also obtain asymptotic behavior of $\eta_2$ in $r\rightarrow0$ and $r\rightarrow+\infty$ from the equation satisfied by $\eta_2$:
\begin{equation}
	\begin{aligned}
		\left. \eta_2\right| _{r\rightarrow0}&\sim r^{\nu-1},\\
		\left. \eta_2\right| _{r\rightarrow+\infty}&\sim \left\lbrace \begin{matrix}
			N_2\frac{I_{\nu+\frac{D}{2}-2}(v_gr)}{r^{\frac{D}{2}-1}},&\text{if}&v\neq0,&\\
			N_2\left( r^{2-D}-\frac{L^2}{2\hat{m}_\phi }r^{1-D}\right) ,&\text{if}&v=0&\& &\hat{m}_\phi\neq0,\\
			N_2r^{\nu-D+1},&\text{if}&v=0&\& &\hat{m}_\phi=0&\&&D>3.
		\end{matrix} \right. 
	\end{aligned}
\end{equation}

Therefore,
\begin{equation}
	\begin{aligned}
		\lim_{r\rightarrow0}\left(I_1 + I_2\right) &=\frac{\nu-1}{\nu+D-3},\\
		\lim_{r\rightarrow+\infty}\left(I_1 + I_2\right) &=\left\lbrace\begin{matrix}
			0,&\text{if}&v\neq0,\\
			\frac{\nu-1}{\nu+D-3},&\text{else}.\\
		\end{matrix} \right. 
	\end{aligned}
\end{equation}

On the other hand, when $v\neq0$, $\det\hat{\Psi}_\nu$ can be obtained directly directly by subjecting $\det\Psi_\nu$ to the following transformation
\begin{equation}
	\begin{aligned}
		\bar{\phi}_g&\rightarrow v_g,\quad \bar{\phi}^\prime_g\rightarrow 0,\\
		\eta_2 &\rightarrow \hat{\eta}_2= \Gamma(\nu+\frac{D}{2}-1)\left(\frac{2}{v_g}\right)^{\nu+\frac{D}{2}-2} \frac{I_{\nu+\frac{D}{2}-2}(v_gr)}{r^{\frac{D}{2}-1}}.
	\end{aligned}
\end{equation}

We obtain
\begin{equation}
	\begin{aligned}
		\det\hat{\Psi}_\nu
		&\sim \frac{r^{2\nu-D+1}\partial_r\left( \hat{\eta}_2r^{D-2}\right)}{\left(2\nu+D-2\right)v_g} \left[\xi+\frac{\left(\nu-1\right)\left(2\nu+D-2\right)}{r^{\nu+2}\partial_r\left( \hat{\eta}_2r^{D-2}\right)}\left( \int_{0}^{r}\hat{\eta}_2r_1^{\nu+D-2} \d r_1\right)\right],\\
		\left. \det\hat{\Psi}_\nu\right| _{r\rightarrow0} 
		&\sim \frac{r^{2\nu-D+1}\partial_r\left( \hat{\eta}_2r^{D-2}\right)}{\left(2\nu+D-2\right)v_g} \left(\xi+\frac{\nu-1}{\nu+D-3}\right),\\
		\left. \det\hat{\Psi}_\nu\right| _{r\rightarrow+\infty} 
		&\sim \frac{r^{2\nu-D+1}\partial_r\left( \hat{\eta}_2r^{D-2}\right)}{\left(2\nu+D-2\right)v_g} \xi. 
	\end{aligned}
\end{equation}

Therefore,
\begin{equation}
	\begin{aligned}
		\lim_{r\rightarrow0}\frac{\det\Psi_\nu}{\det\hat{\Psi}_\nu}&=\frac{v_g}{\phi_{cg}},\\
		\lim_{r\rightarrow+\infty}\frac{\det\Psi_\nu}{\det\hat{\Psi}_\nu}&=\lim_{r\rightarrow+\infty} \frac{v_g\partial_r\left( \eta_2r^{D-2}\right)}{\bar{\phi}_g\partial_r\left( \hat{\eta}_2r^{D-2}\right)}.
	\end{aligned}
\end{equation}

When $v=0$, we should start with 
\begin{equation}
	\begin{aligned}
		\widehat{\mathcal{M}}_{\nu>1}^{Da}&=\begin{pmatrix}
			-\Delta_\nu+\frac{D-1}{r^2}&-\frac{2L}{r^2}&0\\
			-\frac{2L}{r^2}&-\Delta_\nu-\frac{D-3}{r^2}&0\\
			0&0&-\Delta_\nu+\hat{m}_\phi^2
		\end{pmatrix}+\left(1-\frac{1}{\xi}\right)\begin{pmatrix}
			\Delta_2&-L\partial_r\frac{1}{r}&0\\
			\frac{L}{r^D}\partial_rr^{D-1}&-\frac{L^2}{r^2}&0\\
			0&0&0
		\end{pmatrix}.
	\end{aligned}
\end{equation}

We take
\begin{equation}
	\begin{aligned}
		\det\hat{\Psi}_\nu
		&=\left|\begin{matrix}
			\left(\nu-1\right)r^{\nu-2}&\frac{\nu-1-\left(\nu+1\right)\xi}{2L}r^{\nu}&0\\
			Lr^{\nu-2}&\frac{\nu+D-1-\left(\nu+D-3\right)\xi }{2\left(\nu+D-3\right)}r^{\nu}&0\\
			0&0&F_\nu(r;\hat{m}_\phi)
		\end{matrix}\right|=\left( \xi +\frac{\nu-1}{\nu+D-3}\right)F_\nu(r;\hat{m}_\phi)r^{2\nu-2},
	\end{aligned}
\end{equation}
so there are
\begin{equation}
	\begin{aligned}
		\lim_{r\rightarrow0}\frac{\det\Psi_\nu}{\det\hat{\Psi}_\nu}&=\frac{\nu+D-3}{\left(2\nu+D-2\right)\phi_{cg}},\\
		\lim_{r\rightarrow+\infty}\frac{\det\Psi_\nu}{\det\hat{\Psi}_\nu}&=\lim_{r\rightarrow+\infty} \frac{\partial_r\left( \eta_2r^{D-2}\right)}{\left(2\nu+D-2\right) r^{D-3}\bar{\phi}_gF_\nu(r;\hat{m}_\phi)}.
	\end{aligned}
\end{equation}

Thus, we can obtain
\begin{equation}
	\begin{aligned}
		R^{\text{Da}}_{\nu>1}=\lim_{r\rightarrow+\infty} \frac{\phi_{c}\partial_r\left( \eta_2r^{D-2}\right) }{\left(\nu+D-3\right) r^{D-3}\bar{\phi}\hat{\Psi}^{\text{Da}}_\nu(r)}.
	\end{aligned}
\end{equation}

Therefore, if we denote $\Psi^{\text{Da}}_\nu(r)=\frac{\phi_{c}\partial_r\left( \eta_2r^{D-2}\right) }{\left(\nu+D-3\right) r^{D-3}\bar{\phi}}$, $\Psi^{\text{Da}}_\nu(r)$ satisfies the following equation
\begin{equation}
	\begin{aligned}
		\left( \Delta_\nu -m_a^2\right) \Psi^{\text{Da}}_\nu(r)=&\frac{\partial_r\left(\bar{\phi}^2_g r^2\right)}{\bar{\phi}^2_g r^2+L^2}\left(\partial_r -\frac{\bar{\phi}_g^\prime}{\bar{\phi}_g}\right)\Psi^{\text{Da}}_\nu(r).
	\end{aligned}
\end{equation}
\subsection{Summary}
Finally, we finally write $\ln\mathcal{A}_X=\sum\limits_{\nu=1}^{\infty}\text{deg}_X(\nu;D) \ln\left( \frac{\Psi^X_\nu}{\hat{\Psi}^X_\nu}\right).$ Here, $\hat{\Psi}^{X}_\nu(r)$ can be written out explicitly. we obtain
\begin{enumerate}
	\item Higgs
		\begin{equation}
			\hat{\Psi}^{\text{H}}_\nu(r)= F_\nu(r;\hat{m}_\phi),
		\end{equation}
	\item Fermion
		\begin{equation}
			\hat{\Psi}^{\psi}_\nu(r)=F_\nu(r;v_y),
		\end{equation}
	\item Da mode
		\begin{equation}
			\hat{\Psi}^{\text{Da}}_\nu(r)=\left\lbrace \begin{matrix}
				F_\nu(r;v_g)\left(1+\frac{v_gr}{\nu+D-3}\frac{I_{\bar{\nu}+1}(v_gr)}{I_{\bar{\nu}}(v_gr)}\right) ,&\text{if}&v\neq0,\\
				F_\nu(r;\hat{m}_\phi),&\text{if}&v=0,
			\end{matrix}\right. 
		\end{equation}
	\item T mode
		\begin{equation}
			\hat{\Psi}^{\text{T}}_\nu(r)=F_{\nu+1}(r;v_g).
		\end{equation}
\end{enumerate}

In this way, we can later conveniently use the WKB approximation to carry out an expansion for large angular momentum.
\section{Details of WKB Approximation}
As described in the main text, we can compute $C^X_n(r;D)$ by substituting Eq.(23) into Eqs.(15-18). Below are the detailed calculation process and results for each particle.
\subsection{Higgs}
Consider an equation of the form
\begin{equation}
	\Delta_\nu\psi=W\psi.
\end{equation}
Here, $\delta W=m_\phi^2-\hat{m}_\phi^2$ for the Higgs case (and $\delta W=\bar{\phi}^2_g-v_g^2$ for the T mode case).

Setting $\psi=r^{1-\frac{D}{2}}\Psi$ and $r=e^x$, we obtain 
\begin{equation}\label{equ-102}
	\partial_x^2 \Psi=\left( e^{2x}W+\bar{\nu}^2\right) \Psi,
\end{equation}
where $\partial_x=\frac{\d}{\d x}$. By expanding $\partial_x\ln \Psi$ into the form of  $\sum\limits_{n=0}^{+\infty} C_n\bar{\nu}^{1-n}$, and then requiring the coefficients of same powers of $\bar{\nu}$ to be identical, we obtain
\begin{equation}
	\begin{aligned}
		C_0=&1,\quad C_1=0 ,\quad C_2=\frac{1}{2}e^{2x}W,\\
		C_3=&-\frac{1}{4}\partial_x\left( e^{2 x} W\right) ,\\
		C_4=&-\frac{1}{8} e^{4 x}  W^2+\frac{1}{8}\partial_x^2\left(e^{2 x}W\right) ,\\
		C_5=&\frac{1}{8}\partial_x\left( e^{4 x}  W^2\right) -\frac{1}{16} \partial_x^3 \left( e^{2 x}W\right) ,\\
		C_6=&\frac{1}{16} e^{6 x}  W^3-\frac{1}{8} e^{4 x} W^2+\frac{1}{32}e^{4 x}\left( \partial_x W\right)^2+\frac{1}{32}\left( 2-3\partial_x\right) \partial_x\left( e^{4 x}  W^2\right)  +\frac{1}{32} \partial_x^4 \left( e^{2 x}W\right)\\
		C_{m+1}=&-\frac{1}{2}\partial_xC_{m}-\frac{1}{2}\sum_{i=2}^{m-1}C_{i}C_{m+1-i},\quad \text{for}\quad m=3,4,5,\dots
	\end{aligned}
\end{equation}
Since the bounce approaches the false vacuum at infinity, the total differential term does not contribute after integration and can be directly removed. Thus, for $\delta \partial_x\ln \Psi\equiv \partial_x\left(\ln \Psi-\ln \hat{\Psi}\right)$, these yield
\begin{equation}
    \begin{aligned}
    	\delta C_0=&\delta C_1=\delta C_{2m+1}=0\\
		\delta C_2=&\frac{1}{2}e^{2x}\delta W,\\
		\delta C_4=&-\frac{1}{8} e^{4 x} \delta\left( W^2\right) ,\\
		\delta C_6=&\frac{1}{16} e^{6 x} \delta\left( W^3\right)-\frac{1}{8}  e^{4 x} \delta\left( W^2\right)+\frac{1}{32}e^{4 x}\delta\left[ \left( \partial_x W\right)^2\right] .
	\end{aligned}
\end{equation}

For the add-back items, noting that $\bar{\nu}=\frac{\Gamma(\nu+\frac{D}{2}-1)}{\Gamma(\nu+\frac{D}{2}-2)}$ and $\Gamma(x+\nu-1)=\left(x\right)_{\nu-1}\Gamma(x)$, we can see
\begin{equation}
	\text{Deg}_\text{H}(n;D)=\left(\frac{D}{2}-1\right)^{1-n}{}_{n-1}F_{n-2}[D-2,\underbrace{\frac{D}{2}-1,\dots,\frac{D}{2}-1}_{n-2 \text{ terms of } \frac{D}{2}-1};\underbrace{\frac{D}{2},\dots,\frac{D}{2}}_{n-2 \text{ terms of } \frac{D}{2}};1],
\end{equation}
where ${}_{n-1}F_{n-2}[\cdots]$ is generalized hypergeometric function.

It can be seen that the divergent terms are only contained in the terms with $n \leq D$ when $D > 2$ and we can conveniently obtain the divergence of this function in specific D dimensions. 

Finally, in 4 dimensions, the renormalized result is obtained as 
\begin{equation}
    \begin{aligned}
    	\left[\text{Deg}_\text{H}(2;4)\delta C^H_{2}(r;4)\right]_{re} =&0,\\
    	\left[\text{Deg}_\text{H}(4;4)\delta C^H_{4}(r;4)\right]_{re} =&-\frac{1}{8} r^3\delta m_\phi^4\left(1+\gamma_E+\ln\frac{\mu Rr}{2}\right),\\
    	\left[\text{Deg}_\text{H}(6;4)\delta C^H_{6}(r;4)\right]_{re}=&\zeta_R(3)\delta C^H_{6}(r;4),
    \end{aligned}
\end{equation}
where $\zeta_R(x)$ is the Riemann Zeta Function.

In 3 dimensions, the renormalized result is
\begin{equation}
	\begin{aligned}
		\left[\text{Deg}_H(2;3)\delta C^H_{2}(r;3)\right]_{re} =&0,\\
		\left[\text{Deg}_H(4;3)\delta C^H_{4}(r;3)\right]_{re} =&6\zeta_R(2)\delta C^H_{4}(r;3),\\
		\left[\text{Deg}_H(6;3)\delta C^H_{6}(r;3)\right]_{re}=&30\zeta_R(4)\delta C^H_{6}(r;3).
	\end{aligned}
\end{equation}

\subsection{Fermion}
Consider an equation of the form
\begin{equation}
	\Delta_\nu \psi=U\partial_r\psi-\frac{\nu-1}{r}U\psi+W\psi,
\end{equation}
where $\delta W=\delta\bar{\phi}^2_y=\bar{\phi}^2_y-v_y^2$ and $\delta U=\frac{\bar{\phi}^\prime_y}{\bar{\phi}_y}$.

Setting $\psi=r^{1-\frac{D}{2}}\Psi$ and $r=e^x$, we obtain
\begin{equation}\label{equ-2.3.80}
	\partial_x^2 \Psi=e^{x}U\partial_x \Psi+\left( e^{2x}W-\bar{\nu} e^{x}U+\bar{\nu}^2\right) \Psi.
\end{equation}

Thus, $C^\psi_n$ will be0
\begin{equation}
	\begin{aligned}
		&C^\psi_0=1,\quad C^\psi_1=0,\quad C^\psi_2=\frac{1}{2}e^{2x}W,\quad 2C^\psi_{3}=e^{x}UC^\psi_2-\partial_xC^\psi_2,\\
		&C^\psi_{m+1}=\frac{1}{2}e^{x}UC^\psi_{m}-\frac{1}{2}\partial_xC^\psi_{m}-\frac{1}{2}\sum_{i=2}^{m-1}C^\psi_{i}C^\psi_{m+1-i},\quad \text{for}\quad m=3,4,5,\dots\\
	\end{aligned}
\end{equation}

Therefore, noting $e^{x}U=\frac{1}{2}\frac{\partial_x W}{W} $, we obtain
\begin{equation}\label{equ-Cpsi}
	\begin{aligned}
		\delta C^\psi_0(r;D)=&\delta C^\psi_1(r;D)=0,\quad \delta C^\psi_2(r;D)=-2\delta C^\psi_3(r;D)=\frac{1}{2}r\delta\bar{\phi}_y^2 ,\\
		\delta C^\psi_4(r;D)=&\frac{1}{4}r\delta\bar{\phi}_y^2 -\frac{1}{8}r^3\delta\bar{\phi}_y^4-\frac{1}{8}r^3\left( \partial_r\bar{\phi}_y\right) ^2,\\
		\delta C^\psi_5(r;D)=&-\frac{1}{4}r\delta\bar{\phi}_y^2+\frac{3}{16}r^3\delta\bar{\phi}_y^4 +\frac{3}{16}r^3\left( \partial_r\bar{\phi}_y\right) ^2,\\
		\delta C^\psi_6(r;D)=&\frac{1}{4}r\delta\bar{\phi}_y^2-\frac{1}{2}r^3\delta\bar{\phi}_y^4 +\frac{1}{16}r^5\delta\bar{\phi}_y^6-\frac{11}{32}r^3\left( \partial_r\bar{\phi}_y\right)^2+\frac{5}{16}r^5\bar{\phi}_y^2\left( \partial_r\bar{\phi}_y\right)^2+\frac{1}{32}r^5\left( \partial_r^2\bar{\phi}_y\right)^2.\\
	\end{aligned}
\end{equation}

For the add-back items, note that for $2^{\lfloor\frac{D}{2}\rfloor}$, we can always choose an appropriate limiting direction for $\epsilon$ such that it remains unchanged. Therefore, we regard it as a numerical value independent of $\epsilon$. We can see
\begin{equation}
	\begin{aligned}
		\text{Deg}_\psi(n;D)=2^{\lfloor\frac{D}{2}\rfloor}\left(\frac{D}{2}-1\right)^{1-n}
		{}_nF_{n-1}[D-1,\underbrace{\frac{D}{2}-1,\dots,\frac{D}{2}-1}_{n-1 \text{ terms of } \frac{D}{2}-1};\underbrace{\frac{D}{2},\dots,\frac{D}{2}}_{n-1 \text{ terms of } \frac{D}{2}};1].
	\end{aligned}
\end{equation}
The divergent terms are only contained in the terms with $n \leq D$ when $D > 2$.

For comparison with Ref.~\cite{baratellaFunctionalDeterminantsFalse2025}, we choose 
\begin{equation}
	\begin{aligned}
		&\text{Deg}_\psi(2;4)\delta C^\psi_2(r;4)+\text{Deg}_\psi(3;4)\delta C^\psi_3(r;4)+\left(\text{Deg}_\psi(4;4)-2\zeta(2)\right)\delta C^\psi_4(r;4)\\
		=&\sum_{\nu=1}^{+\infty}\left(\nu+\frac{1}{2}\right)r\delta\bar{\phi}_y^2-\frac{1}{4\nu} r^3\left[\delta\bar{\phi}_y^4 +\left( \partial_r\bar{\phi}_y\right) ^2\right]
	\end{aligned}
\end{equation}
as the subtraction terms, then the addback term is
\begin{equation}
	\begin{aligned}
		&\left[\text{Deg}_\psi(2;4)\delta C^\psi_2(r;4)+\text{Deg}_\psi(3;4)\delta C^\psi_3(r;4)+\left(\text{Deg}_\psi(4;4)-2\zeta(2)\right)\delta C^\psi_4(r;4)\right]_{re}\\
		=&-\frac{1}{4}r^3\left[\delta\bar{\phi}_y^4 +\left( \partial_r\bar{\phi}_y\right) ^2\right]\left( 1+\gamma_E+\ln\frac{\mu Rr}{2}\right).
	\end{aligned}
\end{equation}

We can also obtain the addback term corresponding to $\sum\limits_{n=2}^{3}\text{Deg}_\psi(n;D)\delta C^\psi_n(r;3)$ is
\begin{equation}
	\begin{aligned}
		\left[\sum\limits_{n=2}^{3}\text{Deg}_\psi(n;3)\delta C^\psi_n(r;3)\right]_{re}=-\left(3\zeta(2)+2\right) \frac{1}{4}r\delta\bar{\phi}_y^2.
	\end{aligned}
\end{equation}

As a consistency check, we employ the results for $D=4$ in the $\bar{\phi}=\sqrt{\frac{8}{\left|\lambda\right|}}\frac{R}{R^2+r^2}$ to verify $\delta C^\psi_5(r;D)$ and $\delta C^\psi_6(r;D)$ in \eqref{equ-Cpsi}. It can be seen that the $R^{\psi}_\nu$ satisfies:
\begin{align}
	\ln R^{\psi}_\nu=&\ln \left[\frac{\Gamma^2(\nu+1)}{\Gamma(\nu+1-i\frac{y}{\sqrt{\left|\lambda\right|}})\Gamma(\nu+1+i\frac{y}{\sqrt{\left|\lambda\right|}})}\right]\\
	\sim & \frac{y^2}{\left|\lambda\right|}\nu^{-1}-\frac{y^2}{2\left|\lambda\right|}\nu^{-2}-\frac{\left(y^2-\left|\lambda\right|\right) y^2}{6\lambda^2 }\nu^{-3}+\frac{y^4}{4 \lambda^2}\nu^{-4}+\frac{\left(2 y^4-5 \left|\lambda\right| y^2-\lambda^2\right) y^2}{30\left|\lambda\right|^3 }\nu^{-5},
\end{align}
and the computed values from the \eqref{equ-Cpsi} give $\delta C^\psi_5(r;D)=\frac{y^4}{4 \lambda^2}$ and $\delta C^\psi_6(r;D)=\frac{\left(2 y^4-5 \left|\lambda\right| y^2-\lambda^2\right) y^2}{30\left|\lambda\right|^3 }$, which are consistent with that relation.

\subsection{Guage boson}
Consider an equation of the form
\begin{equation}
	\left[A+\bar{\nu}^2-\left(\frac{D}{2}-1\right)^2\right] \left( \Delta_\nu \psi-W\psi \right)  =\partial_rA\left(\partial_r\psi-B\psi\right).
\end{equation}

Following the same procedure, we obtain (denote $\frac{D}{2}-1\equiv l$)
\begin{equation}
	\begin{aligned}
		&\left(A+\bar{\nu}^2-l^2\right) \left[  \partial_x^2 \Psi-\left(\bar{\nu}^2+ e^{2x}W\right) \Psi \right] =\partial_xA\left[\partial_x\Psi-\left( l+e^{x}B\right) \Psi\right],\\
		&C^{\text{Da}}_0=1,\quad C^{\text{Da}}_1=0,\quad C^{\text{Da}}_2=\frac{1}{2}e^{2x}W,\quad C^{\text{Da}}_{3}=\frac{1}{2}\partial_xA-\frac{1}{2}\partial_xC^{\text{Da}}_2,\\
		&C^{\text{Da}}_{4}=-\frac{1}{8}e^{4x}W^2-\frac{1}{2}\left(l+e^xB\right) \partial_xA-\frac{1}{2}\partial_xC^{\text{Da}}_3,\\
		&D^{\text{Da}}_{m}=C^{\text{Da}}_{m-3}\partial_xA-\left(A-l^2\right)D^{\text{Da}}_{m-3} ,\quad\text{for}\quad m=5,6,7\dots\\
	\end{aligned}
\end{equation}
where $D^{\text{Da}}_{m}\equiv\partial_xC^{\text{Da}}_{m-1}+2C^{\text{Da}}_{m}+\sum\limits_{i=2}^{m-2}C^{\text{Da}}_{i}C^{\text{Da}}_{m-i}$.

Thus, for the equations of Da mode of the guage boson, we obtain
\begin{equation}\label{equ-2.3.92}
	\begin{aligned}
		\delta C^{\text{Da}}_0(r;D)=&\delta C^{\text{Da}}_1(r;D)=\delta C^{\text{Da}}_{2m+1}(r;D)=0,\\
		\delta C^{\text{Da}}_2(r;D)=&\frac{1}{2}r\delta m_a^2,\\
		\delta C^{\text{Da}}_4(r;D)=&r\delta \bar{\phi}_g^2 -\frac{1}{8}r^3\delta m_a^4-r^3\left(\partial_r\bar{\phi}_g\right)^2 ,\\
		\delta C^{\text{Da}}_6(r;D)=&\frac{\left(D-2\right)^2}{4}r\delta \bar{\phi}_g^2-\frac{3}{2}r^3\delta \bar{\phi}_g^4-\frac{\left(D-2\right)^2}{4}r^3\left(\partial_r\bar{\phi}_g\right)^2 +\frac{3}{2}r^5\bar{\phi}_g^2\left(\partial_r\bar{\phi}_g\right)^2-\frac{1}{8}r^3\delta m_a^4 +\frac{1}{16}r^5\delta m_a^6 \\
		&+\frac{1}{32}r^5\left[ \partial_r\left( m_a^2\right)\right] ^2 +\frac{D-12}{4}r^3\left( m_a^2\bar{\phi}_g^2-v_g^4\right)+\frac{1}{4}r^5m_a^2\bar{\phi}_g^2\left(m_a^2-\bar{\phi}_g^2\right)-\frac{3}{4}r^4 \bar{\phi}_g^2\partial_r\left( m_a^2\right)\\
		&+\frac{3}{4}r^5 m_a^2\left( \partial_r\bar{\phi}_g\right) ^2.
	\end{aligned}
\end{equation}

We know $\text{deg}_{\text{Da}}(\nu;D)=\text{deg}_{\text{H}}(\nu;D)$. Therefore, $\text{Deg}_{\text{Da}}(n;D)=\text{Deg}_\text{H}(n;D)$. However, it should be noted that in higher dimensions, $\delta C^{\text{Da}}_n(r;D)$ itself depends on $D$ and thus also requires expansion.

For the case of $R^{\text{T}}_{\nu}$, it can be observed that if $\nu+1$ is transformed into $\nu$, this scenario becomes equivalent to the case discussed for the Higgs. Therefore, we can directly derive the result for this scenario by reverting $\nu$ back to $\nu+1$ in the Higgs result. The result obtained is
\begin{equation}\label{equ-2.3.94}
	\begin{aligned}
		\delta C^{\text{T}}_2(r;D)=&\frac{1}{2}r\delta \bar{\phi}_g^2 ,\\
		\delta C^{\text{T}}_4(r;D)=&-\frac{1}{8}r^3\delta \bar{\phi}_g^4,\\
		\delta C^{\text{T}}_6(r;D)=&\frac{1}{16}r^5\delta \bar{\phi}_g^6 -\frac{1}{8}r^3\delta \bar{\phi}_g^4 +\frac{1}{8}r^5\bar{\phi}^2_g\left(\partial_r\bar{\phi}_g\right)^2.
	\end{aligned}
\end{equation}

For the addback term in this case, we obtain
\begin{equation}
	\begin{aligned}
		\text{Deg}_{\text{T}}(n;D)=&\left(D-1\right)\left(\frac{2}{D}\right)^{n-2}{}_{n+1}F_{n}[2,D,D-2,\underbrace{\frac{D}{2},\dots,\frac{D}{2}}_{n-2 \text{ terms of } \frac{D}{2}};3,D-1,\underbrace{\frac{D}{2}+1,\dots,\frac{D}{2}+1}_{n-2 \text{ terms of } \frac{D}{2}+1};1],
	\end{aligned}
\end{equation}
and the divergent terms are only contained in the terms with $n \leq D$ when $D > 2$.

We choose 
\begin{equation}
	\begin{aligned}
		&\sum_{\nu=1}^{+\infty}\frac{1}{2}r\left[\nu\delta m_a^2 +2\left(\nu+1\right)\delta \bar{\phi}_g^2 \right]-\frac{1}{8\nu}r^3\left[\delta m_a^4+2\delta \bar{\phi}_g^4+8 \left(\partial_r\bar{\phi}_g\right)^2\right]\\
		=&\text{Deg}_{\text{Da}}(2;4)\Delta C^{\text{Da}}_2(r;4)+\text{Deg}_{\text{Da}}(4,4)\Delta C^{\text{Da}}_4(r;4)+\left[\text{Deg}_\text{T}(2;4)-\sum_{\nu=1}^{+\infty}\frac{2}{\nu\left(\nu+1\right)}\right]\Delta C^\text{T}_2(r;4)\\
		&+\left[\text{Deg}_\text{T}(4;4)+\sum_{\nu=1}^{+\infty}\frac{2}{\nu\left(\nu+1\right)}+\sum_{\nu=1}^{+\infty}\frac{2}{\left(\nu+1\right)^3}\right]\Delta C^\text{T}_4(r;4)
	\end{aligned}
\end{equation}
as the subtraction terms for comparison with Ref.~\cite{baratellaFunctionalDeterminantsFalse2025}, then the addback term is
\begin{equation}
	\begin{aligned}
		\frac{1}{2}r\delta \bar{\phi}_g^2+\frac{1}{8}r^3\delta \bar{\phi}_g^4-\frac{1}{8}r^3\left[\delta m_a^4+2\delta \bar{\phi}_g^4+8 \left(\partial_r\bar{\phi}_g\right)^2\right]\left(1+\gamma_E+\ln\frac{\mu Rr}{2}\right).
	\end{aligned}
\end{equation}

For $D=3$, we can obtain
\begin{equation}
	\begin{aligned}
		\left[\text{Deg}_\text{Da}(2;3)\delta C^\text{Da}_{2}(r;3)+\text{Deg}_\text{T}(2;3)\delta C^\text{T}_{2}(r;3)\right]_{re} =&-2\delta C^\text{T}_{2}(r;3),\\
		\left[\text{Deg}_\text{Da}(4;3)\delta C^\text{Da}_{4}(r;3)+\text{Deg}_\text{T}(4;3)\delta C^\text{T}_4(r;3)\right]_{re} =&6\zeta_R(2)\delta C^\text{Da}_{4}(r;3)+\left( 6\zeta_R(2)-8\right) \delta C^\text{T}_4(r;3),\\
		\left[\text{Deg}_\text{Da}(6;3)\delta C^\text{Da}_{6}(r;3)+\text{Deg}_\text{T}(6;3)\delta C^\text{T}_6(r;3)\right]_{re}=&30\zeta_R(4)\delta C^\text{Da}_{6}(r;3)+(30\zeta_R(4)-32)\delta C^\text{T}_6(r;3).
	\end{aligned}
\end{equation}

For \eqref{equ-2.3.92}, we carried out the same verification as in the fermionic part and found agreement.

\section{Numerical Computation Aspects}
The target equations exhibits a divergence at $r=0$. It is therefore necessary to perform Taylor expansions of the functions $\bar{\phi}(r)$ and $\Psi_\nu^{X}(r)$ in order to avoid the singularity in the equations. The Taylor expansion of the function $\bar{\phi}(r)$ is
\begin{equation}
	\bar{\phi}(r)=\phi_c + \phi_2 r^2 + \phi_4 r^4 + \phi_6 r^6+ O( r^8),
\end{equation}
where
\begin{equation}
	\begin{aligned}
		U_i&\equiv \left. \frac{\d^iU}{\d \bar{\phi}^i}\right|_{\bar{\phi}=\phi_c},\quad\phi_2=\frac{1}{2D}U_1,\quad\phi_4=\frac{1}{8D\left(D+2\right)}U_1U_2,\\
		\phi_6&=\frac{1}{48D\left(D+4\right)}\left(\frac{1}{D+2}U_2^2+\frac{1}{D}U_1U_3\right)U_1.
	\end{aligned}
\end{equation}

And the Taylor expansions of the $\Psi_\nu^{X}(r)$'s are
\begin{enumerate}
	\item Higgs
	\begin{equation}
		\left. \Psi_\nu^{\text{H}}(r)\right| _{r\rightarrow0}\sim r^{\nu-1}\left[ 1+\frac{U_2}{2\left(2\nu+D-2\right)}r^2+\frac{1}{8\left( 2\nu+D\right)}\left(\frac{1}{2\nu+D-2}U_2^2+\frac{1}{D}U_1U_3\right)r^4 +O( r^{6})\right],
	\end{equation}
	\item Fermion
	\begin{equation}
		\left. \Psi^\psi_\nu(r)\right| _{r\rightarrow0}\sim r^{\nu-1}\left[ 1+\frac{\phi_{cy}^2}{2\left(2\nu+D-2\right)}r^2+\frac{\phi_{cy}^4+4\left(2\nu+D-1\right)\phi_{cy}\phi_{2y}}{8\left(2\nu+D-2\right)\left(2\nu+D\right)}r^4+O(r^6) \right],
	\end{equation}
	\item Da mode
	\begin{equation}
		\begin{aligned}
			\left.\Psi_\nu^{\text{Da}}(r)\right|_{r\rightarrow+0}\sim& r^{\nu-1}\left\lbrace1+\left[D\frac{\phi_{2y}}{\phi_{cy}}+\frac{\nu+D-1}{2\left(\nu+D-3\right)}\phi_{cy}^2\right]\frac{r^2}{2\nu+D-2}+\left[\left(D+2\right)\frac{\phi_{4y}}{\phi_{cy}}-\frac{D\left(\nu-1\right)}{\left(2\nu+D-2\right)}\frac{\phi_{2y}^2}{\phi_{cy}^2}\right.\right.\\
			&\left.\left.+\frac{\left(\nu+D+1\right)\phi_{cy}^4+8\left(\nu^2+2D\nu+D^2-D-3\right)\phi_{cy}\phi_{2y}}{8\left(2\nu+D-2\right)\left(\nu+D-3\right)}\right]\frac{r^4}{2\nu+D}+O(r^6)\right\rbrace,
		\end{aligned}
	\end{equation}
	\item T mode
	\begin{equation}
		\left. \Psi_\nu^{\text{T}}(r)\right| _{r\rightarrow0}\sim r^{\nu}\left[1+\frac{\phi_{cg}^2}{2\left(2\nu+D\right)}r^2+\frac{\phi_{cg}^4+4\left(2\nu+D\right)\phi_{cg}\phi_{2g}}{8\left(2\nu+D\right)\left(2\nu+D+2\right)}r^4+O(r^6)\right].
	\end{equation}
\end{enumerate}

As mentioned in the main text, the equations we actually perform numerical calculations on are
\begin{equation}\label{equ-119}
	\begin{aligned}
		\partial_r^2R^\text{H}_\nu(r)=&\left( \delta m_\phi^2\right)  R^\text{H}_\nu(r)-\left(2\frac{\partial_r \hat{\Psi}_\nu^{\text{H}}(r)}{\hat{\Psi}_\nu^{\text{H}}(r)}+\frac{D-1}{r}\right) \partial_rR^\text{H}_\nu(r),\\
		\partial_r^2R^\psi_\nu(r)=&\left[ \delta \bar{\phi}_y^2+\frac{\bar{\phi}_y^\prime}{\bar{\phi}_y}\left(\frac{\partial_r \hat{\Psi}_\nu^\psi(r)}{\hat{\Psi}_\nu^\psi(r)}-\frac{\nu-1}{r}\right) \right] R^\psi_\nu(r)-\left(2\frac{\partial_r \hat{\Psi}_\nu^\psi(r)}{\hat{\Psi}_\nu^\psi(r)}+\frac{D-1}{r}-\frac{\bar{\phi}_y^\prime}{\bar{\phi}_y}\right) \partial_rR^\psi_\nu(r),\\
		\partial_r^2R_\nu^{\text{Da}}(r)=&\left\lbrace  \delta m^2_a+\left[ \frac{\partial_r\left(\bar{\phi}^2_g r^2\right)}{\bar{\phi}^2_g r^2+L^2}-\frac{2v^2_g r}{v^2_g r^2+L^2}\right] \frac{\partial_r \hat{\Psi}_\nu^\text{Da}(r)}{\hat{\Psi}_\nu^\text{Da}(r)}-\frac{\bar{\phi}_g^\prime}{\bar{\phi}_g}\frac{\partial_r\left(\bar{\phi}^2_g r^2\right)}{\bar{\phi}^2_g r^2+L^2}\right\rbrace R_\nu^{\text{Da}}(r) \\
		&-\left[2\frac{\partial_r \hat{\Psi}_\nu^\text{Da}(r)}{\hat{\Psi}_\nu^\text{Da}(r)}+\frac{D-1}{r}-\frac{\partial_r\left(\bar{\phi}^2_g r^2\right)}{\bar{\phi}^2_g r^2+L^2}\right]\partial_r R^\text{Da}_\nu(r),\\
		\partial_r^2R^\text{T}_\nu(r)=&\left( \delta \bar{\phi}_g^2\right)  R^\text{T}_\nu(r)-\left(2\frac{\partial_r \hat{\Psi}_\nu^{\text{T}}(r)}{\hat{\Psi}_\nu^{\text{T}}(r)}+\frac{D-1}{r}\right) \partial_rR^\text{T}_\nu(r).\\
	\end{aligned}
\end{equation}
Furthermore, for the Higgs sector where $\hat{m}_\phi=0$, noting that $\frac{\d^2U}{\d \bar{\phi}^2}\sim O(r^{-2\left(D-2\right)})$ when $r\rightarrow+\infty$, we have
\begin{equation}
	R^\text{H}_\nu(r)\simeq R^\text{H}_\nu+Cr^{-2\nu-D+4},\quad\text{if}\quad D>3.
\end{equation}

Therefore, $R^\text{H}_\nu(r)+\frac{r\partial_rR^\text{H}_\nu(r)}{2\nu+D-4}$ and $R^\text{H}_\nu(r)$ share the same limit, with $R^\text{H}_\nu(r)+\frac{r\partial_rR^\text{H}_\nu(r)}{2\nu+D-4}$ converging more rapidly. We redefine $R^\text{H}_\nu(r)+\frac{r\partial_rR^\text{H}_\nu(r)}{2\nu+D-4}\rightarrow R^\text{H}_\nu(r)$ in this case.

Likewise, when $v=0$ and $\hat{m}_\phi>0$, or when $v=\hat{m}_\phi=0$ and $D>3$, the same pattern can be observed in the T mode sector, allowing us to redefine $R^\text{T}_\nu(r)+\frac{r\partial_rR^\text{T}_\nu(r)}{2\nu+D-2}\rightarrow R^\text{T}_\nu(r)$.
{\color{blue}
\begin{figure}[!h]
	\centering
	\begin{minipage}{0.45\textwidth}
		\centering
		\includegraphics[width=\linewidth]{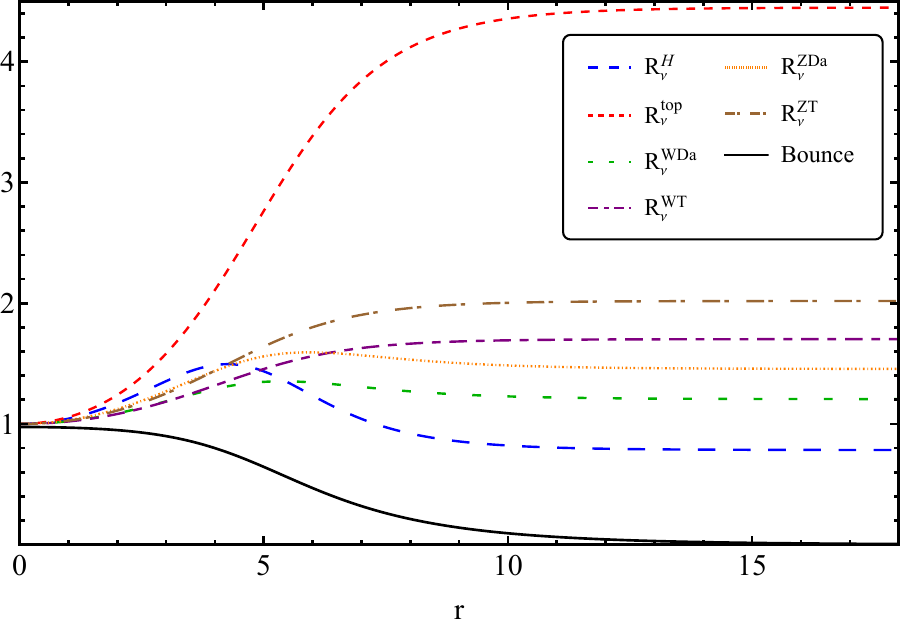}
		\caption{Convergence of the equations($\nu=7$).}
		\label{fig-1}
	\end{minipage}
	\hfill
	\begin{minipage}{0.45\textwidth}
		\centering
		\includegraphics[width=\linewidth]{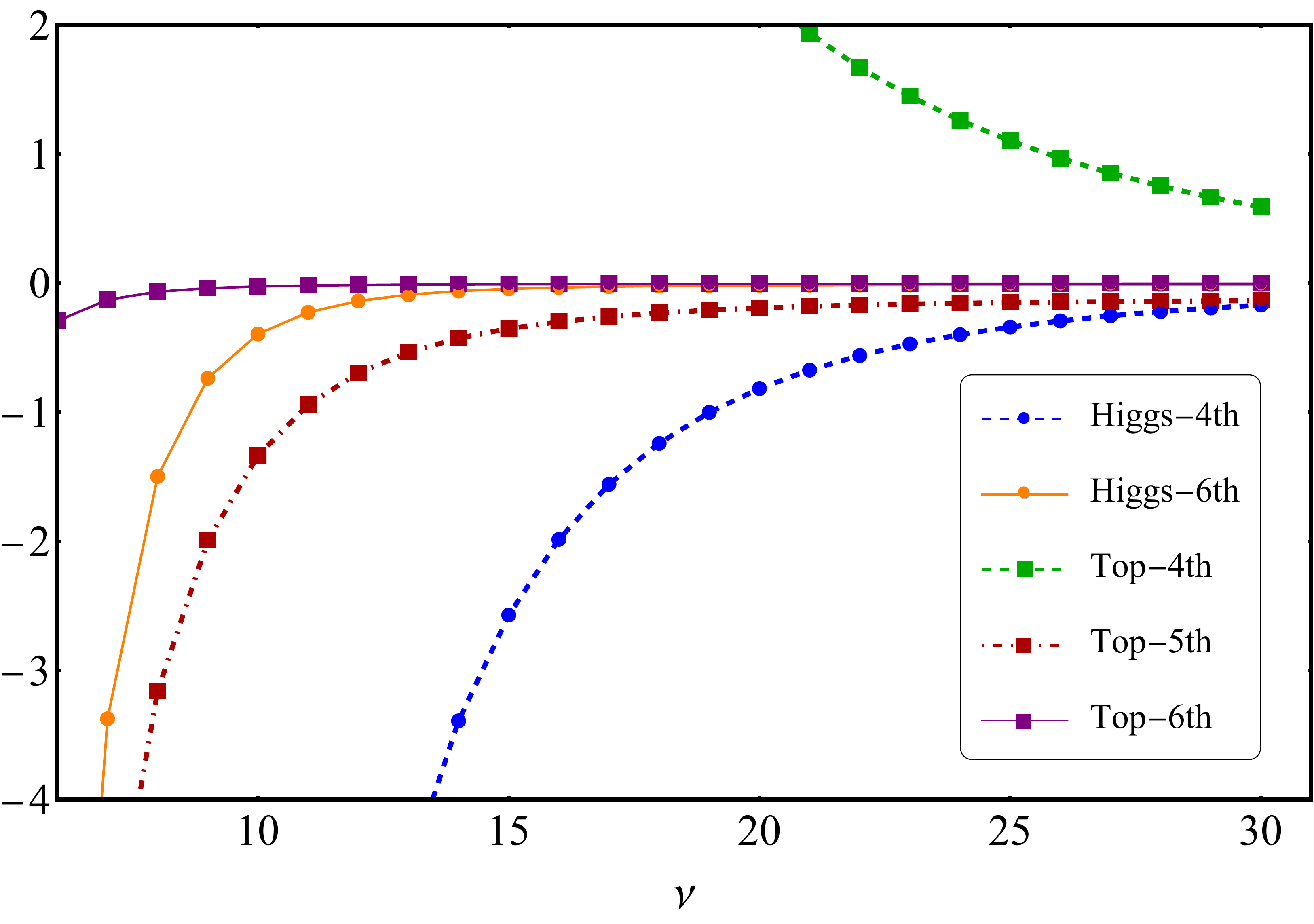}
		\caption{Residuals of the WKB approximation for Higgs($\times10^4$) and top quarks($\times10^3$).}
		\label{fig-2}
	\end{minipage}
\end{figure}
}
Figure \ref{fig-1} and Figure \ref{fig-2} provide supplementary details for the $D=4$ example in the main text. Figure \ref{fig-1} shows the convergence of the different equations in \eqref{equ-119}. It can be noted that each equation converges to its limiting value extremely rapidly, at a rate comparable to that of the bounce solution. This is especially notable when compared with the equations for bosons presented in Ref.~\cite{endoFalseVacuumDecay2017} — the latter have a convergence radius of approximately $\frac{L^2}{2\hat{m}_\phi}$, which is disastrous for numerical calculations at large angular momenta. In contrast, the equations in this work always exhibit convergence rates similar to that of the bounce, regardless of the magnitude of the angular momentum. Figure \ref{fig-2} shows the effect of the WKB approximation for other particles.

\bibliography{reference}